\documentclass[aps,physrev,reprint,groupedaddress]{revtex4-2}

\usepackage[utf8]{inputenc}
\usepackage{amsmath, amssymb}
\numberwithin{equation}{section} 
\usepackage{algorithm}
\usepackage{algorithmic}
\usepackage{fancyhdr} 
\usepackage{lipsum} 
\usepackage{setspace} 
\usepackage[]{fncychap} 
\usepackage{xcolor} 
\usepackage{hyperref}
\usepackage{tabularx}
\usepackage{appendix}
\usepackage{textcomp}
\usepackage{verbatim}
\usepackage{mathtools}
\usepackage[a4paper, total={6.5in, 8.5in}]{geometry}
\NewDocumentCommand{\evalat}{sO{\big}mm}{%
  \IfBooleanTF{#1}
   {\mleft. #3 \mright|_{#4}}
   {#3#2|_{#4}}%
}
\usepackage{cleveref}
\usepackage{graphicx}
\usepackage{textalpha}
\usepackage{silence}
\newlength{\saveparindent}
\newlength{\saveparskip}
\AtBeginDocument{%
  \setlength{\saveparindent}{\parindent}%
  \setlength{\saveparskip}{\parskip}%
}

\newcommand{\C}{\mathbb{C}}
\newcommand{\plas}{\boldsymbol{\varepsilon}^\textrm{p}}
\newcommand{\elas}{\boldsymbol{\varepsilon}^\textrm{e}}

\usepackage{dcolumn}
\usepackage{bm}

\begin{document}

\preprint{APS/123-QED}

\title{Bridging Atomistic and Continuum Descriptions of Nanoscale Dislocation Loops in Tungsten} 

\author{Joseph Duque Lopez$^1$}
 \email{joseph.duque@warwick.ac.uk}
\author{Sergei Dudarev$^2$}
\author{James Kermode$^1$}
\author{Thomas Hudson$^3$}%
\affiliation{$^1$WCPM, School of Engineering, University of Warwick, CV4 7AL, Coventry, United Kingdom.}
\affiliation{$^2$CCFE, United Kingdom Atomic Energy Authority, Culham Science Centre, Abingdon, Oxfordshire, OX14 3EB, United Kingdom.}
\affiliation{$^3$Warwick Mathematics Institute, University of Warwick, CV4 7AL, Coventry, United Kingdom.}

\date{\today}

\begin{abstract}
In order to predict the long-term effects of irradiation on the material properties of tungsten, a continuum approach to simulating the interactions of dislocation loops, which arise from radiation damage, is proposed. Continuum models of the displacement, strain and stress fields produced by dislocation loops exhibit unphysical singularities near the defect core, but are thought to accurately capture atomistic displacements in the far-field. A linear elastic model of nanoscale dislocation loops in tungsten is developed, and the model is verified using atomistic simulations to ensure that the model is informed by lower-length scale phenomena such that the physics of the problem is correctly captured. We discuss the model and its advantages, and show that predictions produced by atomistic simulations do indeed agree well with the far-field behaviour of the continuum model when dislocation loops are far from material boundaries. In particular, we robustly demonstrate that the decay rate of atomistic results and continuum results coincide with one another, and show that the results converge as the size of the atomistic simulations approach the far-field limit. 
\end{abstract}

\maketitle

\section{Introduction}
In the field of nuclear fusion, tungsten is considered the primary choice as an armour material for plasma-facing components in a fusion reactor \cite{barabash_armour_1999, agency_iter_2002, bolt_materials_2004, kaufmann_tungsten_2007, nygren_making_2011}. Reasons for this include the fact that tungsten has the highest melting point of any known natural metal, has a low physical sputtering yield, exhibits low thermal expansion and good thermal conductivity \cite{lopez-cazalilla_effect_2023, gilbert_neutron-induced_2011}. Such properties are desirable in armour components to maximise their erosion lifetimes when subjected to irradiation.
Tungsten's relatively short neutron activated half-life also makes it ideal for exposure to radiation because it can be more frequently recycled \cite{seidel_activation_2004}, although the timescale for this phenomena is still in the order of $\sim$ 100 years. However, prolonged exposure to radiation inevitably causes the physical and mechanical properties of a material to degrade over time \cite{yoshida_review_1999, barabash_neutron_2000, hu_irradiation_2016, fellman_radiation_2019, PWM2020}.
In order to study the effects of irradiation damage in tungsten, a range of atomistic simulations have been performed in recent years \cite{PWM2020, fikar_molecular_2009, srivastava_dislocation_2013, swinburne_theory_2013, kobayashi_molecular_2015, swinburne_fast_2016, de_backer_multiscale_2018, xu_molecular_2024, das_dislocation_2024}, with many employing multiscale approaches and interatomic potentials informed by density functional theory (DFT) data. Atomistic simulations are used to study defects because the use of more accurate, quantum-scale DFT codes is limited by the computational cost of the accessible spatial scales, which ranges to a few thousand atoms at most \cite{dudarev_chapter_2025}.

While simulations using interatomic potentials provide many useful insights beyond what can be achieved with DFT calculations, they nevertheless remain limited to short time- and length-scale phenomena \cite{uberuaga_computational_2020, ruiz_pestana_chapter_2023}, particularly when compared to the size of fusion plasma-facing components and their expected operational lifetimes. As an example, plasma-facing materials are expected to operate for around 2.5 full-power years \cite{toschi_how_2001} with surface areas of around 850 \(\text{m}^2\) \cite{merola_iter_2010} over a 5-year replacement cycle \cite{bolt_materials_2004}, clearly well beyond the time- and length-scales which are achievable with molecular dynamics. As a bridge towards modelling such components, it is therefore necessary to work with continuum models to enable prediction of the material evolution under irradiation over long periods of time. Continuum models are useful because they can model relatively large-scale and long-time phenomena when compared to atomistic simulations, trading detail for computational efficiency. Such models must nevertheless be informed by observations of the microscopic phenomena that occur during irradiation to verify that the predicted quantities are accurate.

In order to model radiation damage at the continuum level, the underlying physics must therefore be understood first. At the atomistic level, radiation damage in plasma-facing materials can be explained as the accumulation of Frenkel pair defects \cite{thompson_damage_1960, fitzgerald_shape_2009, fitzgerald_structure_2018}, alongside clusters of point defects which are formed from high-energy collision cascades \cite{fu_effect_2017, PWM2020}. These clusters then relax into more energetically favourable configurations, such as dislocation loops and voids \cite{Fitzgerald2018, sand_high-energy_2013, sand_radiation_2014, sand_cascade_2017, wang_dynamic_2023}. As the number of dislocations grows they begin to interact and form complex structures \cite{rovelli_statistical_2018, li_diffusion_2019, wang_dynamic_2023}. In order to predict how such dislocation networks interact in the mesoscale regime --- with length-scales on the order of $10^{-7} \, \text{m}$ --- one must model how a given dislocation interacts with a given stress field. Classical dislocation theory predicts that dislocations create plastic strains in a bulk material, and induce stress fields leading to long-range elastic interactions with other dislocations. This interaction force is characterised by the Peach-Koehler formula \cite{peach_forces_1950,Bulatov2020,Anderson2017}.
Thanks to the existence of exact solutions to these problems in the context of continuum linear elastoplasticity (CLE) theory, continuum models have been used to study dislocations since the early 20th century, and remain a successful way to predict the behaviour of such defects at a macroscopic scale \cite{Bacon1980,balluffi_introduction_2012,Anderson2017,Bulatov2020}. Although the underlying assumption of a continuous material breaks down close to defects, even relatively close to the dislocation cores, continuum theories can still be effective in studying material response \cite{Blin1955}.

Motivated by the challenge of studying defect interaction at the mesoscale in irradiated materials, in this paper, we investigate the accuracy of CLE predictions for the atomistic displacements around a dislocation loop.
Thanks to the weight of existing evidence, CLE theory is expected to predict the far-field displacements around individual dislocation loops. Our aim is to elucidate more precisely at what distance away from the defect the predictions between CLE and such simulations agree, and therefore establishing a separation length-scale at which CLE theory can be used to make predictions of macroscopic properties of defect ensembles. To this end, an asymptotic description of the displacement, stress and strain fields far from a self interstitial atom (SIA) dislocation loop are calculated using the CLE model in which the loop is represented as a force dipole, and atomistic simulations are performed to verify the continuum predictions against the displacements in tungsten using a range of appropriate interatomic potentials. In our investigation, we find that atomistic simulations agree with the displacement field predicted by the CLE theory at distances of around twice the loop radii away from the dislocation loop centre for a range of different dislocation loop radii.

Due to the need to truncate atomistic simulations, the verification of the CLE prediction in an infinite medium using atomistic simulations is subtle. Care needs to be taken both in choosing boundary conditions, and finite size effects remain significant even at relatively large system sizes. Our analysis shows convergence in the dislocation loop area as the overall simulation size increases, supporting our modelling choice to represent dislocation loops as a CLE force dipole.

The remainder of the paper is structured as follows: in \cref{sec:CLEmodel}, we outline the CLE model for a dislocation loop. In \cref{sec:atomistic_method}, we describe the methodology used to generate and relax atomistic structures. The results obtained are discussed in \cref{sec:results}, and we conclude in \cref{sec:conclusion}.
\section{CLE prediction around a dislocation loop}
\label{sec:CLEmodel}
\subsection{The Volterra equation}
\begin{figure}[hbpt!]
    \centering
    \includegraphics[width=0.9\linewidth]{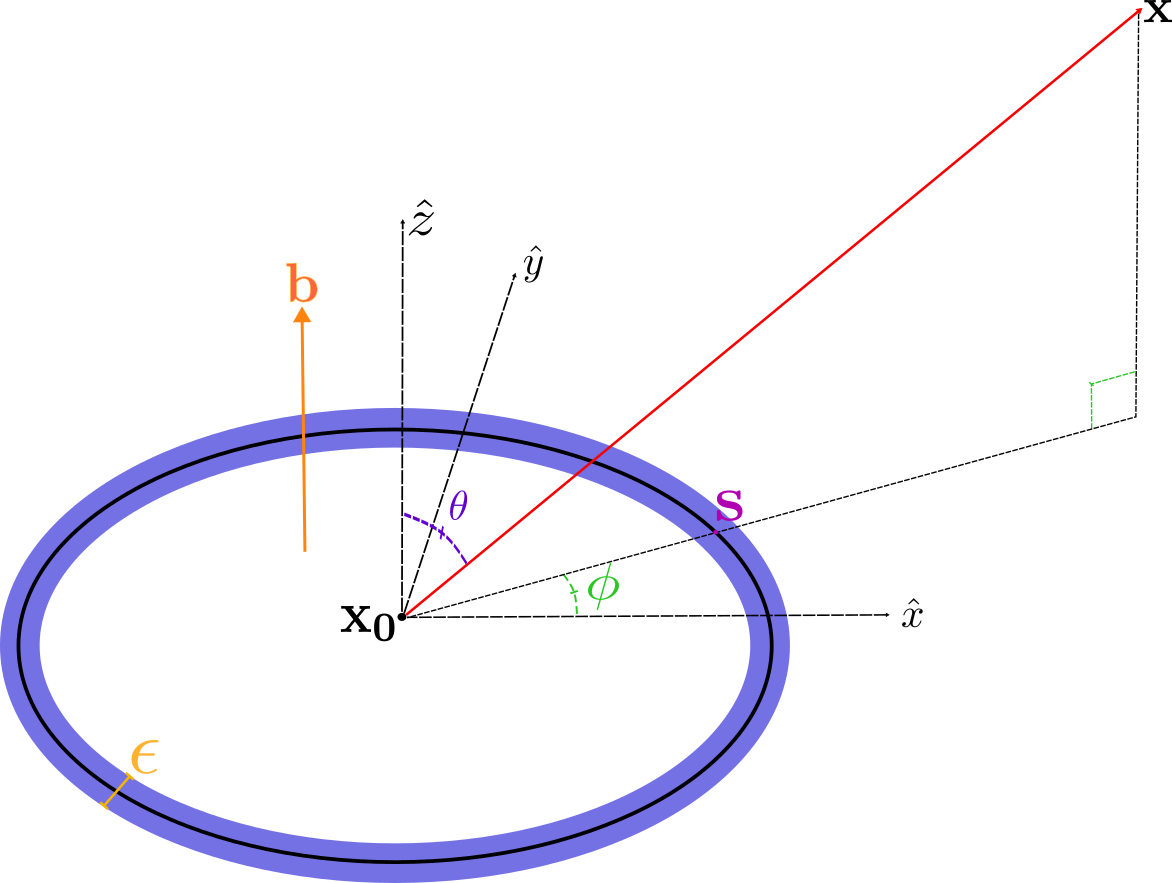}
    \caption{A dislocation loop $\mathbf{s}$, with a Burgers vector $\mathbf{b}$ and loop centre $\mathbf{x_{0}}$. $\mathbb{\epsilon}$ indicates a dislocation core radius, indicating the region within which continuum theory breaks down.}
    \label{fig:dislocation-loop-setup}
\end{figure}

Classical dislocation theory begins by considering the force balance equation of static CLE theory and making some key assumptions about the strain fields involved. Doing so, one can obtain an expression for the displacement field \(\mathbf{u}\) at some point \(\mathbf{x}\) due to a dislocation lying along a line contour \(\mathbf{s}\) in the material; \cref{fig:dislocation-loop-setup} illustrates the basic geometry of the problem we consider.

As usual in linear elastoplastic theory, the total strain is assumed to be additively decomposed into elastic and plastic contributions, \(\boldsymbol{\varepsilon} = \elas + \plas\) \cite{Mura1982, balluffi_introduction_2012}, and
Saint-Venant's compatibility condition \cite{n_i_muskhelishvili_basic_1948} is assumed to apply for the total strain.
Manipulations using the elastic Green's function allow us to arrive at an integral representation of the displacement, known as either the Mura or Volterra equation \cite{Bacon1980, Mura1982, cai_non-singular_2006, clouet_dislocation_2009, yin_computing_2012, gao_displacement_2015, dudarev_elastic_2017}:
\begin{equation}
\label{eq:volterraEq}
    u_i(\mathbf{x}) = \int_{\Sigma} \C_{jklm} \, b_{l} \, n_m (\mathbf{x}')\, G_{ij,k}(\mathbf{x} - \mathbf{x}') \,  \, \mathrm{d} S(\mathbf{x}').
\end{equation}
Here, we apply the Einstein summation convention, and indices appearing after a comma refer to differentiation with respect to the relevant Cartesian coordinate of position. The surface $\Sigma$ is any arbitrary surface with consistently oriented boundary curve $\mathbf{s}$. \(G_{ij}(\mathbf{x})\) is the elastic Green's function, a tensor field which encodes the $i$th component of displacement along the \(\mathbf{x}\) direction in response to a unit point force applied in $j$th coordinate direction at the origin \cite{bacon_anisotropic_1980, balluffi_introduction_2012}. $\C$ is the elasticity tensor of the material, \(b_l\) is the component of the Burgers vector in the \(l\)-direction, and \(n_m\) is the $m$th component of the unit normal to $\Sigma$. The primed position variable \(\mathbf{x'}\) is an integration point on the surface $\Sigma$. 

\subsection{Asymptotics for far-field displacements}

Since we wish to predict the behaviour far from the dislocation loop, we now consider the relationships between the characteristic length-scales in this problem. Taking $R$ to the be radius of the dislocation loop, $\epsilon$ to be the dislocation core radius and $r=|\mathbf{x}-\mathbf{x}_0|$ to be a typical distance to the dislocation core centre at which we wish to make a prediction, we will assume that $\epsilon\ll R\ll r$. In this regime, the displacement predicted by \cref{eq:volterraEq} can be approximated by treating each dislocation loop as a point-like force dipole in the medium. Assuming a perfectly circular dislocation loop of radius \(R\), the equation becomes \cite{dudarev_elastic_2017}: 
\begin{equation}
    u_i(\mathbf{x}) \approx \pi R^2 \, \C_{jklm} \, b_l \,  n_m \, G_{ij,k}(\mathbf{x}-\mathbf{x}_0).
    \label{eq:leading-order-equation}
\end{equation}
One can obtain a higher order terms of the model by carrying out a higher-order asymptotic expansion in \cref{eq:volterraEq}. In the case of a perfectly circular loop lying in the $x_1,x_2$ plane, this yields the approximation
\begin{equation}
\begin{aligned}
    &u_{i}(\mathbf{x}) \approx \pi R^2 \, \C_{jkl3} \, b_l \,  G_{ij,k} (\mathbf{x} - \mathbf{x_0}) 
    \\
    &\; + \frac{\pi R^4}{8} \, \C_{jkl3} \, b_l\, \left( G_{ij, k 1 1} + G_{ij, k 2 2} \right) (\mathbf{x} - \mathbf{x_0}),
\end{aligned}
\label{eq:regularisedVolterra2}
\end{equation}
where the normal vector has been replaced by the unit vector in the $z$-direction, $\mathbf{e}_3$.
It should be noted that recent rigorous results suggest that there may be contributions from nonlinear theory which provide better predictions in the far-field than further expanding the CLE prediction \cite{braun_higher-order_2024}. We therefore choose to truncate the prediction at leading order, using the approximation \cref{eq:leading-order-equation} to make predictions about far-field displacements, strains and stresses.

Noting the dependence of the prediction \cref{eq:leading-order-equation} on the area of the circular loop $\pi R^2$ to leading order, \cref{eq:leading-order-equation} shows that the displacement field is proportional to the area enclosed by the dislocation loop. Moreover, when the Burgers vector $\mathbf{b}$ is parallel to the loop normal $\mathbf{n}$, the entire prefactor can be viewed as being proportional to the added interstitial volume; this is exactly the case of a glide loop induced by the accumulation of SIAs. On the atomistic scale, such dislocation loops cannot immediately collapse, as this requires a local volumetric change achieved through the diffusion of SIAs away from the loop. On the other hand, when the Burgers vector is in the plane, a climb loop is formed. For loops of this type at the scales of interest, we would expect a rapid collapse and annihilation, but the far-field formula discussed above is nevertheless applicable for either glide or climb loops.

\subsection{Discussion}
\label{sec:consequences-model}
Using \cref{eq:volterraEq} to calculate the displacement fields around dislocation loops provides a very general technique for obtaining far-field predictions which is widely used in discrete dislocation (DD) simulations \cite{Bulatov2020}. Much of the computational cost required for DD arises due to the need to evaluate integrals along dislocation lines; the representation \cref{eq:leading-order-equation} however allow us to simply compute the gradient of the Green's function at a single point, providing a more computationally efficient approach suited to studying large ensembles of dislocation loops.

Moreover, we can use \cref{eq:leading-order-equation} to obtain predictions of the elastic strain and hence the stress from this approximation. Using an additive splitting of the total strain, we note that since $\elas = \boldsymbol{\varepsilon}-\plas$ with $\plas$ being concentrated on the surface $\Sigma$ spanning the dislocation loop, the total strain $\boldsymbol{\varepsilon}$ is identical to the elastic strain in the far-field regime. This allows for the computation of the Cauchy stress from Hooke's law, using the total strain $\boldsymbol{\varepsilon}$ in place of $\elas$. An important comment about this approach is that the elastic strain $\elas$ in the far field regime is in fact independent of the surface used to compute the observable; this is a consequence of Stokes' theorem \cite{Bacon1980}. However, the effect of the additional volume needed to create the dislocation loop leads to the displacement field depending explicitly on the excess volume needed to create the dislocation loop, and the plane in which it lies. This information is what is ultimately encoded in the second moment force dipole:
\begin{equation}
    P_{jk} =\C_{jklm}b_ln_m * A_{\text{Loop}}
\end{equation}
where $A_{\text{Loop}}$ is the area of the dislocation loop.

Due to the equation's dependence on the elastic Green's function we see that the total strain behaves like the second derivative of the elastic Green's function. Since the elastic Green's function decays as $|\mathbf{x}|^{-1}$, as one gets further away from the dislocation loop the displacement field $\mathbf{u}$ decays as $r^{-2}$, and the strain and stress should thus decay as $r^{-3}$ due to the linear dependence on $\nabla \mathbf{u}$. This asymptotic decay is identical to that classically predicted for point defects: this connection is unsurprising in retrospect since the prismatic dislocation loops modelled are formed by an accumulation of SIAs.

Our asymptotic result predicts displacements in the far-field regime, and is therefore only valid as long as dislocation loops remain far enough apart. \cite{wang_dynamic_2023} provides a good baseline for the distances in which dislocation loops will interact at. If one thinks of dislocation loops as point particles interacting with one another in a large system, the average interaction distance can be found as \(\langle \boldsymbol{r} \rangle \sim \rho^{-1/3}\) for a density \(\rho\). For radiation damage levels of 0.1 displacements-per-atom (dpa), the density of dislocation loops in tungsten is reported to be around \(2.8\times10^{22} \,\, \text{m}^{-3}\) \cite{wang_dynamic_2023}; this number has been selected as it reflects the highest density in any of the radiation damage regimes studied. In this case, dislocation loops sit at an average spacing of around 330~\AA. Additionally, it is reported that the majority of dislocation loops in this case have a dislocation loop radius of between 25~\AA\ and 50~\AA, so in order to ignore dislocation core effects the distances at which the model should coincide with the atomistic simulations should be around 250~\AA. Consequently, in order to test the validity of the mesoscale approximation derived here a comparison with atomistic simulations at an appropriate far field regime was pursued. The behaviour of the fields predicted far from the defect core was of interest, because dislocation loops are expected to interact at distances which are large multiples of the lattice spacing, as reported in \cite{wang_dynamic_2023}.
\section{Methodology of Atomistic verification}
\label{sec:atomistic_method}

To assess the validity of the displacement prediction obtained in \cref{eq:leading-order-equation} for nanoscale dislocation loops in tungsten, we next compare atomistic simulations with the asymptotic CLE prediction \cref{eq:leading-order-equation}. Simulations were performed where the atomic displacements and approximate strain fields around a simple dislocation loop were calculated from atomistic simulations and compared against the analytical predictions. Convergence between the atomistic simulations and the analytical results is of special interest, to verify that \cref{eq:leading-order-equation} holds in the far-field regime. Code to reproduce the results obtained below is available in an accompanying zenodo repository, see \cite{repo}. 

In tungsten, the most commonly reported dislocation loops observed in irradiated samples are those with Burgers vectors and normals in the geometrically equivalent \(\tfrac12\langle 1 1 1 \rangle \) directions \cite{rau_vacancy_1968, haussermann_elektronenmikroskopische_1972, fukuzumi_defect_2005, chen_new_2018, fitzgerald_structure_2018, das_recent_2019}. Our focus for the verification process will therefore be on dislocation loops of this type, specifically prismatic dislocation loops with \(\tfrac{1}{2}[1 1 1]\) Burgers vectors.

\subsection{Atomistic calculations}
\label{sec:atomistic-calcs}
To set up the atomistic simulations bulk tungsten structures were built using the Atomistic Simulation Environment (ASE) library \cite{HjorthLarsen2017}. The lattice structure used was the body centred cubic (BCC), and the lattice spacing \(a_0\) was obtained by relaxing  a unit cell of tungsten with periodic boundary conditions in LAMMPS \cite{Plimpton1995}. All simulation relaxations were performed with molecular statics, i.e. at $0$ Kelvin.

The primary potential used for all the simulations was an embedded atom model (EAM) from the NIST interatomic potentials repository  \cite{Becker2013, Hale2018, LucasHale2016} by Chen et. al \cite{chen_new_2018}, denoted in this paper as `YC'. This potential was chosen because it predicted the relative stability of different interstitial dislocation loops in W in accordance with experiments. 
In particular, the potential was reported to favour \( \tfrac12 \langle 1 1 1 \rangle \) loops over \( \langle 1 0 0 \rangle \) loops as the loop size increased, in agreement with experimental results. The potential has been used to study molecular dynamic simulations of W radiation damage, for example in \cite{fu_molecular_2019} the potential was used to study the formation of \( \tfrac{1}{2} \langle 1 1 1 \rangle \) interstitial loops during high-energy collision cascades, where it was reported that \( \tfrac{1}{2} \langle 1 1 1 \rangle \) loops dominated the presence of interstitial loops. However, in order to check consistency, a range of different potentials for W-W interactions were also used and compared against the model.

The optimisation algorithm used for most structure relaxations was the Polak-Ribiere conjugate gradient method \cite{polyak_conjugate_1969}, and the stopping criterion for most relaxations was a force tolerance of $10^{-6} \,\, \text{eV}/$\AA. 
For the largest systems, the conjugate gradient method was unable to find the global minimum configuration due to the size of the system and the forces involved. For such relaxations the systems were relaxed as much as possible with the conjugate gradient method, and then a further relaxation on the resulting structures was performed using the FIRE algorithm \cite{bitzek_structural_2006} to ensure that the largest configurations were relaxed below the desired force tolerance.

\begin{figure}[hbt!]
    \centering
    \includegraphics[width=0.5\textwidth]{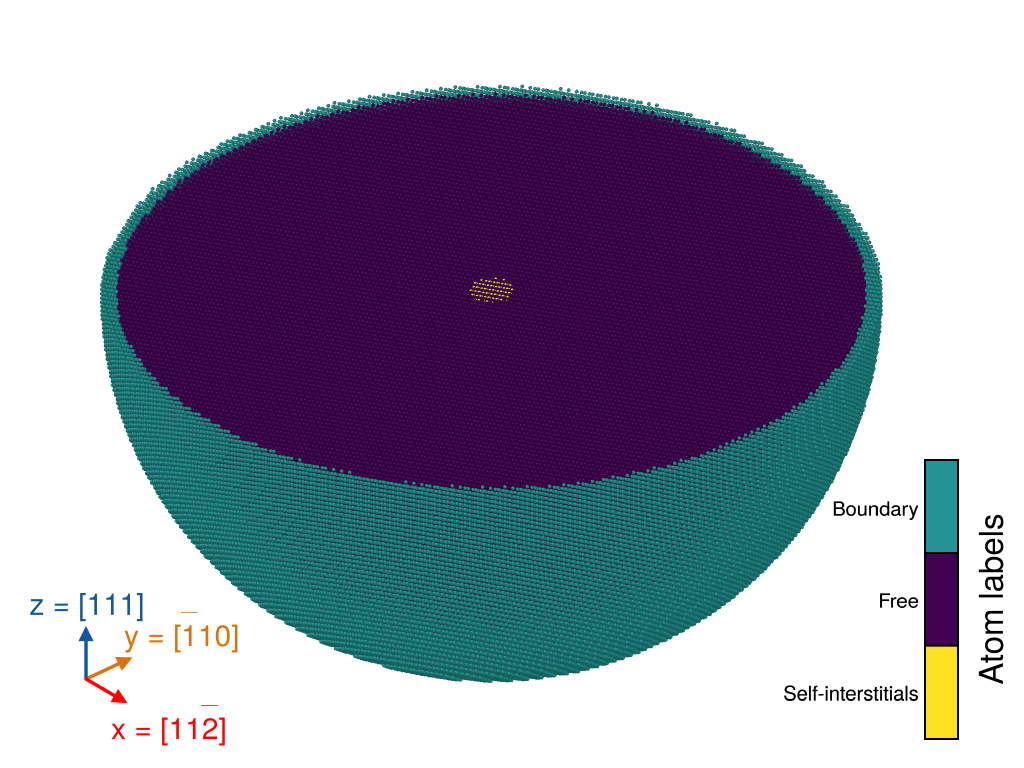}
    \caption{An overview of the atomistic simulation setup. Interstitial atoms were placed at the crowdion sites of a BCC tungsten structure to create a planar dislocation loop in centre of the sphere; the atoms in the boundary region were clamped in position during relaxation.}
    \label{fig:atomistic-setup}
\end{figure}

A schematic of the system simulated can be seen in \cref{fig:atomistic-setup}, which was obtained using the visualisation software OVITO \cite{ovito}. As seen in the figure, the crystallographic orientation of the structure was designed so that the normal vector to the $(111)$ plane was parallel to the $z$-axis. This allowed for intuitive constructions of dislocation loops with a Burgers vector and normal vector that pointed in the \(\tfrac12\langle 1 1 1 \rangle \) direction. As an initial guess for the relaxation, discs of interstitial atoms were placed at the crowdion sites of the BCC structure, as these are the most energetically stable site, as discussed in \cite{paneth_mechanism_1950, derlet_multiscale_2007, fitzgerald_structure_2018}. In addition, the simulation structure was constructed so that the centre of mass of the interstitial atoms coincided with the centre of the bulk sphere. For comparison later, this point is taken to be exactly the point \(\mathbf{x_0}\) defined in \cref{eq:leading-order-equation}. Spherical bulk structures were used in order to properly study the orientational dependence of the elastic fields due to the dislocation loop inserted. The structure was then relaxed in order to obtain atomistic data on the displacement of the atoms due to the introduction of an interstitial dislocation loop.

For the relaxation of the structure, three different boundary conditions on the surface of the sphere were tested, two with fixed displacements applied to atoms in the outer shell, and a traction-free variant in which all atoms were free to relax. For the fixed boundary conditions, a shell of atoms around the sphere was constrained. In the first of these cases, the atoms were constrained to remain at their original crystallographic positions; in the second, the atoms were first displaced according to the continuum prediction \cref{eq:leading-order-equation}. In both cases, the thickness of the clamped boundary region was chosen such that it was larger than the cutoff radius of the EAM potentials used, so that those atoms free to relax but close to the boundary were not affected by the vacuum outside the boundary. For the traction-free variant, the atoms in the outer shell were also allowed to relax freely. In principle, all of these choices of boundary conditions should yield equivalent displacements through a thermodynamic limit procedure as the radius of the system tends to infinity.

The boundary conditions chosen were preferred over periodic boundary conditions in order to avoid image effects from other dislocation loops, and because the spherical nature of the setup conflicts with the choice of a cubic periodic cell. However, the boundary conditions chosen still produce spurious forces --- and hence displacements --- near the boundary. This meant that points near the boundary could not considered to be in the far-field regime due to finite nature of the simulations. 
The effect of the finite size was an important factor to consider because the results obtained should only be compared to the analytic results in the linear elastic regime. Atoms too close to the dislocation loop are subject to discrete core effects which are unable to be accounted for in continuum models, even if the boundary condition at the core is regularised. Additionally, atoms too close to the surface of the sphere are affected by surface effects which the CLE prediction \cref{eq:leading-order-equation} does not take into account.
For this reason the results obtained were only compared for atoms in a central shell, denoted a `region of confidence', between the dislocation core and the sphere boundary. Such region was determined as the region where the boundary effects from both the dislocation core and the surface were minimal, as discussed in \cref{sec:boundary-cond}.

\subsection{Computing continuum predictions}

For comparison with the atomistic results, the continuum prediction fields arising from \eqref{eq:leading-order-equation} were computed. As \cref{eq:leading-order-equation} states, predictions of the displacement field can be obtained at any position $\mathbf{x}$, as long as the Burger's vector, the normal vector, the dislocation loop radius and the elasticity tensor of the material are known. Tungsten has cubic symmetry, so the only Voigt elastic moduli required to construct the full elasticity tensor are \(\C_{11}, \,\, \C_{12}, \,\, \text{and} \,\, \C_{44}\). These moduli were obtained by relaxing a unit cell of tungsten with the chosen potential and extracting the elastic coefficients with the \texttt{matscipy} library \cite{Grigorev2024}. 

Once all the parameters were defined, the displacement field due to the dislocation loop was constructed by looping over all the positions and calculating the displacement of each atom to leading order as defined by the model. The gradient of the elastic Green's function present in \cref{eq:leading-order-equation} was computed using the formulae discussed in \cite{barnett_precise_1972, bacon_anisotropic_1980}. From this, an analytical prediction for the strain and stress fields could be obtained. To compute these robustly, automatic differentiation was employed via the python library JAX \cite{jax2018github}, allowing for efficient computation of $\nabla\mathbf{u}$.
From the displacement gradient, the infinitesimal strain tensor was calculated as \(\boldsymbol{\varepsilon} = \text{sym}(\nabla \mathbf{u})\). As discussed above, in the far-field, the infinitesimal total strain is equal to the elastic strain $\elas$, and so Hooke's law can be used to obtrain the stresses from $\boldsymbol{\varepsilon}$. 

In order to compare the results obtained from the continuum model with the atomistic calculation, the analytic data produced for each atom was saved in extended \texttt{.xyz} files for post-processing, enabling a robust comparison of the CLE strain prediction with approximate strains computed from the atomistic configuration. 

\subsection{Comparing atomistic results and continuum predictions}
\label{sec:comparison_method}
To compare the atomistic and analytic data in a robust manner, and in view of the boundary effects induced by the boundary conditions, multiple atomistic simulations were performed with varying simulation cell size.
To this end, a range of simulation sphere radii were studied, between 50 \AA\ and 500 \AA, with a fixed boundary condition of an outer shell of clamped atoms with a thickness of around 25 \AA, more than twice the cutoff distance of the EAM potential discussed \cite{chen_new_2018}. The largest size simulation of 500 \AA\ was chosen because it provided a big enough region in which to verify the results with the computational resources available.

At these sizes, the results were then compared in a region of the simulation sphere that is far enough away from both the dislocation loop and the simulation surface boundary. Increasing the simulation size increased the region in which the CLE prediction is expected to be valid and where results could be confidently compared. Therefore, as the simulation size was increased, convergence of the atomistic displacements was expected, and therefore increasing agreement between the results from the atomistic simulations and the continuum prediction over larger regions of the simulation. Comparisons were made in multiple ways: qualitatively by looking at the strain fields predicted by both atomistic and continuum theories, and quantitatively by analysing the displacement decay rates on the results, and by demonstrating convergence between the dislocation loop areas between the atomistic and analytic theories.
\section{Results}
\label{sec:results}
We began by investigating the rate of decay of the stresses predicted by the model, in the purely continuum regime, checking that the rate of decay is directly proportional to \(\sim r^{-3}\), the rate of decay of the gradient of displacement field.

Then, we compare the results from the continuum and atomistic perspectives. First, we look at the qualitative pattern in the strain fields predicted by both theories and verify that the atomistic model correctly characterises the field in accordance to the atomistic simulations. 

Next, we looked at how the rate of decay of the displacement fields agree, for the different boundary conditions chosen. Additionally, the analytic results were compared against a range of different interatomic potentials and shown to mostly agree. The effect of the simulation size on the results was also studied to demonstrate the importance of comparing the results in an appropriate far-field regime. 

Finally, the effect of the dislocation loop size was investigated and shown to produce a scaling law in the atomistic simulations, which could be attributed to the finite size effects. These results were then shown to converge to the expected area of a dislocation loop as the simulation size increased.

\subsection{Results from continuum model}

As discussed in \cref{sec:consequences-model}, the analytic model predicts that the stress decays as $r^{-3}$. This rapid algebraic decay results in reasonable far-field approximations of the stress fields which could be used to model interactions between loops when they are sufficiently well--separated.

\begin{figure*}[htpt!]
	\centering
	\centering
	\includegraphics[width=\textwidth]{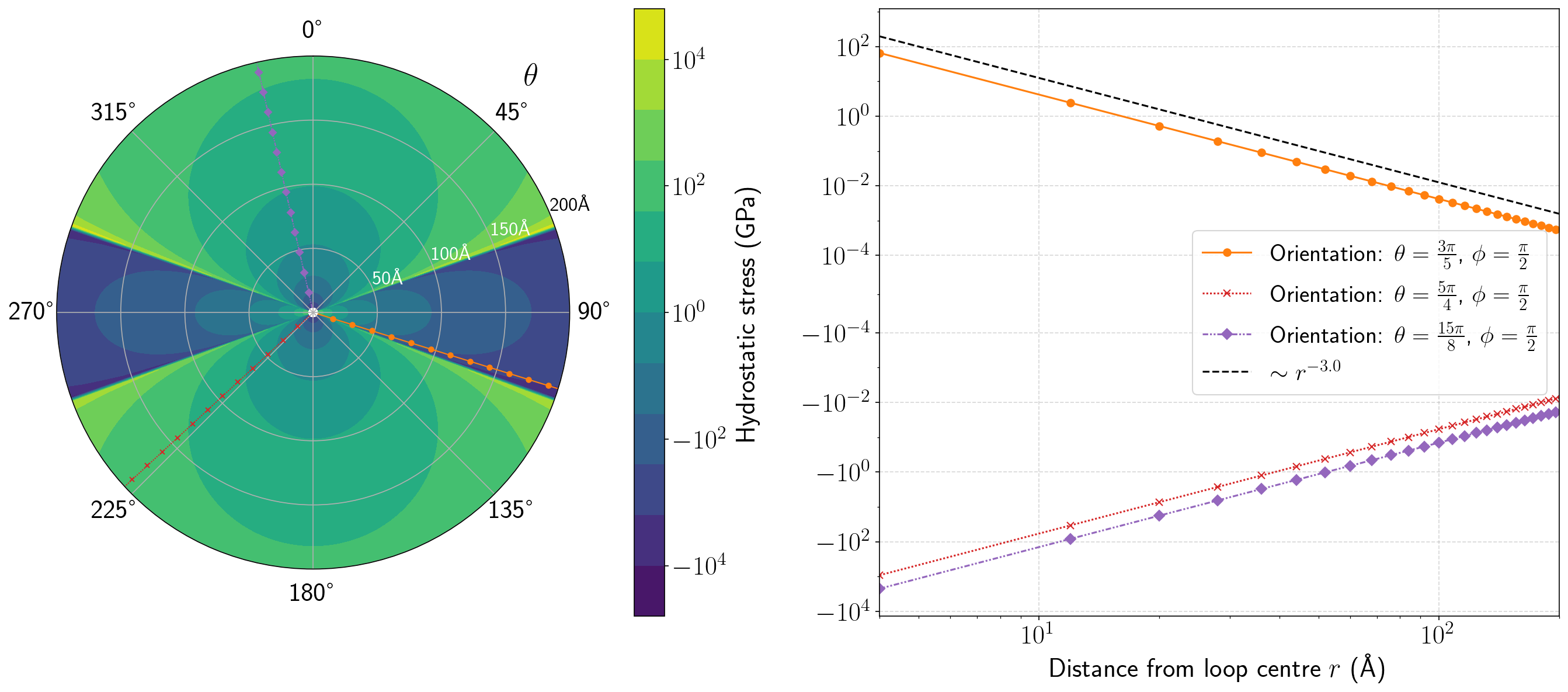}
	\caption{Left: A contour plot of the hydrostatic stress predicted from \cref{eq:leading-order-equation}. A slice is shown along the $y,z$ plane, demonstrating the axisymmetry of the field. Right: Hydrostatic stress predictions along the rays shown, demonstrating the $\sim r^{-3}$ decay rate of field; this decay is identical, irrespective of the orientation.}
	\label{fig:rate_of_decay_contour_volumetric}
\end{figure*}

\begin{figure*}
	\centering
	\includegraphics[width=\textwidth]{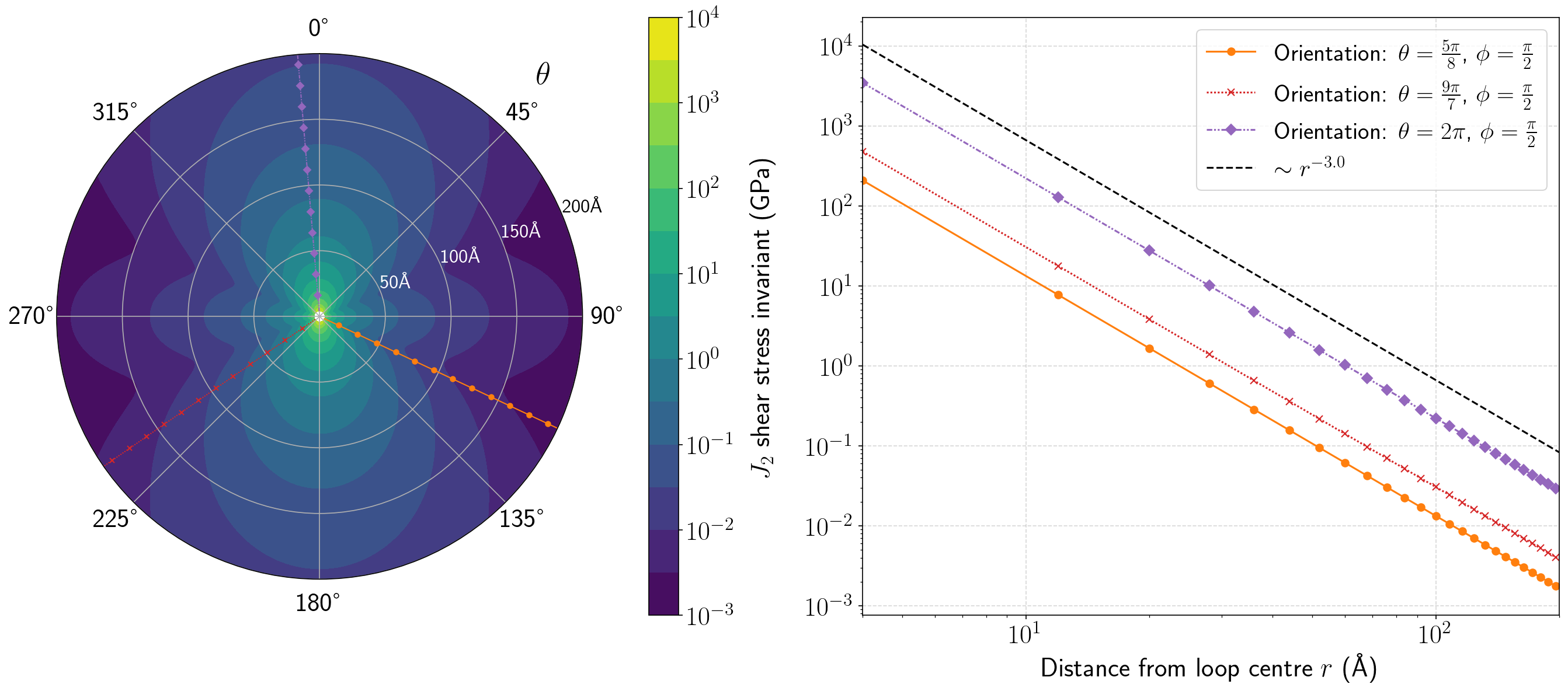}
	\caption{
		Left: A contour plot of the $J_2$ shear stress, predicted from \cref{eq:leading-order-equation}. A slice is shown along the $y,z$ plane, again demonstrating the axisymmetry of the field. Right: $J_2$ stress predictions along the rays shown, demonstrating the $\sim r^{-3}$ decay rate of field.}
	\label{fig:rate_of_decay_contour_J2}
\end{figure*}

\Cref{fig:rate_of_decay_contour_volumetric,fig:rate_of_decay_contour_J2} show the rate of decay of the hydrostatic and shear components of the stress by plotting the stress against the radial distance away from the dislocation for multiple orientations. In these plots the Burgers vector and the normal vector of the dislocation loop were chosen to point in the positive $z$ direction. \Cref{fig:rate_of_decay_contour_volumetric} shows the hydrostatic pressure, $p=-\frac{1}{3}\text{tr}(\mathbf{S})$, obtained as the trace of the predicted elastic stress \(\mathbf{S}\). \Cref{fig:rate_of_decay_contour_J2} shows the $J_2$ shear invariant of the stress tensor. The $J_2$ invariant is defined as 
\begin{equation}
	J_2 = \tfrac16[(\sigma_1 - \sigma_2)^2 + (\sigma_2 - \sigma_3)^2 + (\sigma_3 - \sigma_1)^2],
\end{equation}
where $\sigma_i$ are principal stresses at a given point. The $J_2$ invariant provides a measure of the shear stresses that drive plastic deformation, ignoring the effects of hydrostatic stress. A range of orientations were selected to verify that the rate of decay predicted by the numerical implementation was $r^{-3}$ as expected. For the orientations shown, the results indeed corroborate that the rate of decay is proportional to $r^{-3}$ as expected.

Analysing the continuum stress prediction in other planes demonstrated that the fields are essentially cylindrically symmetric about the line passing through the centre of the dislocation loop normal to the plane on which it lies. This is despite the assumed anisotropy of the elasticity tensor, and likely reflects the very low Zener anisotropy ratio for tungsten which is reported to be around 1.0 \cite{bolef_elastic_1962,lowrie_singlecrystal_1967,mei_elastic_2023}.
The figure also demonstrates that the continuum model loses validity as we approach the dislocation loop: the magnitude of the stresses predicted is on the order of $\sim 10^{3} \,\,\, \text{GPa}$ close to the core, for distances less than the dislocation loop radius of 12 \AA\ chosen. These values are much larger than the experimental bulk and shear moduli of W, which are around 160 GPa and 310 GPa, and hence close to the core the stresses predicted are not physical. 

\subsection{Atomistic Verification}
Using the methodology discussed in \cref{sec:atomistic_method}, atomistic simulations produced relaxed tungsten structures containing dislocation loops. The dislocation loop induced stress fields which led to a displacement of atoms around the dislocation loop. From the displacement fields analysis on the difference between the displacements, and therefore the strain fields, of the atomistic and analytic results was conducted. 

\subsubsection{Strain field comparisons}
An initial analysis into the verification of the model was a comparison on the strain fields that are predicted by the atomistic and continuum theories. In order to calculate the strain fields from an atomistic perspective, functionality within OVITO was used to determine the deformation gradient tensor $\mathbf{F}$ from the deformed and reference configurations \cite{OvitoAtomicStrain}. The equivalent infinitesimal strain $\boldsymbol{\varepsilon}$ was then obtained from $\mathbf{F}$.
From this infinitesimal strain tensor, two scalar quantities were then obtained: the hydrostatic strain and the von Mises local shear strain \cite{OvitoAtomicStrain, li_least-square_nodate}.
These measures of strain were compared with the equivalent scalar quantities obtained from the continuum strain computations.

\begin{figure*}[hbpt!] 
	\centering
	\includegraphics[width=0.33\textwidth]{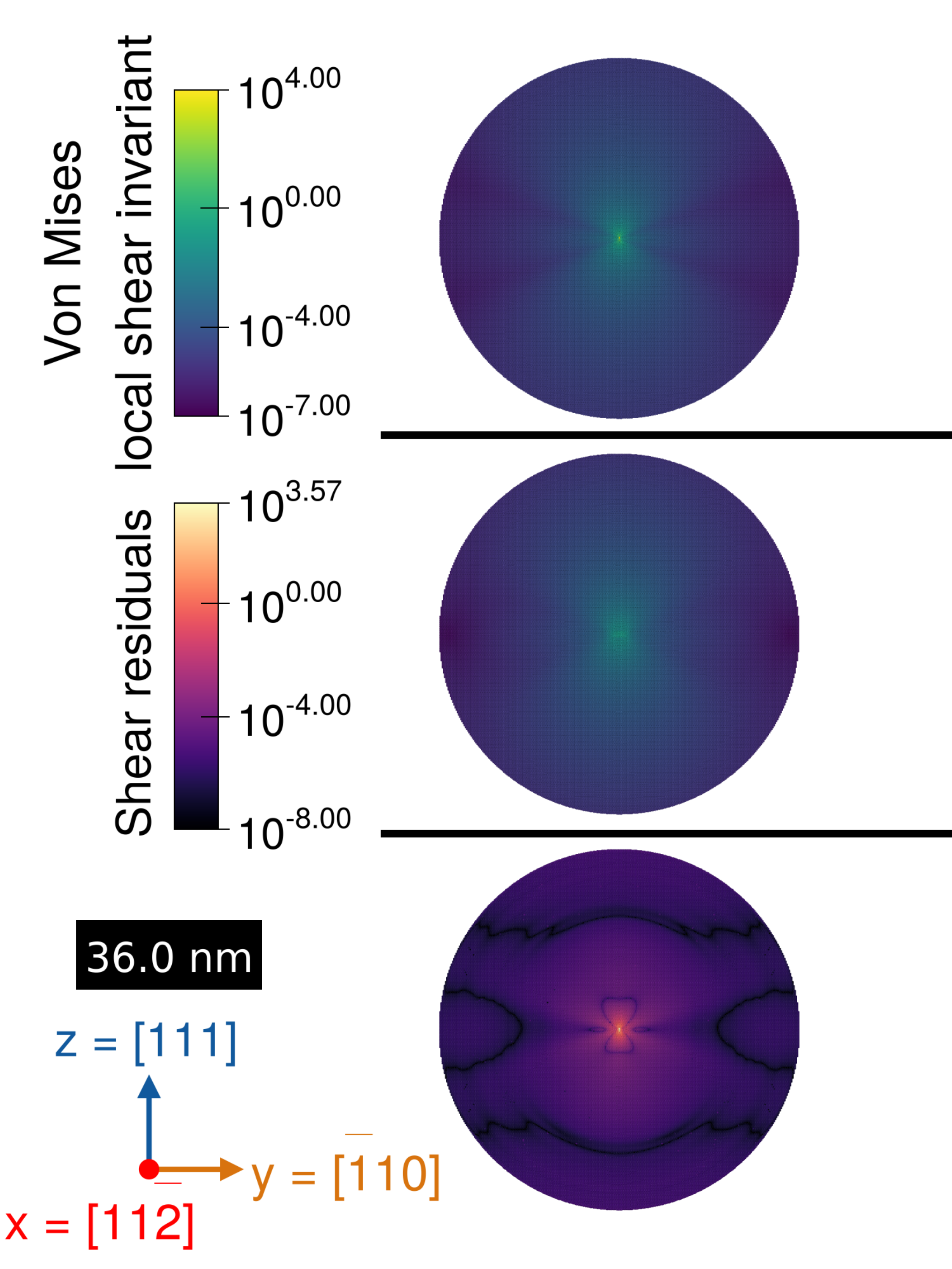}\hspace{-2mm}
	\includegraphics[width=0.33\textwidth]{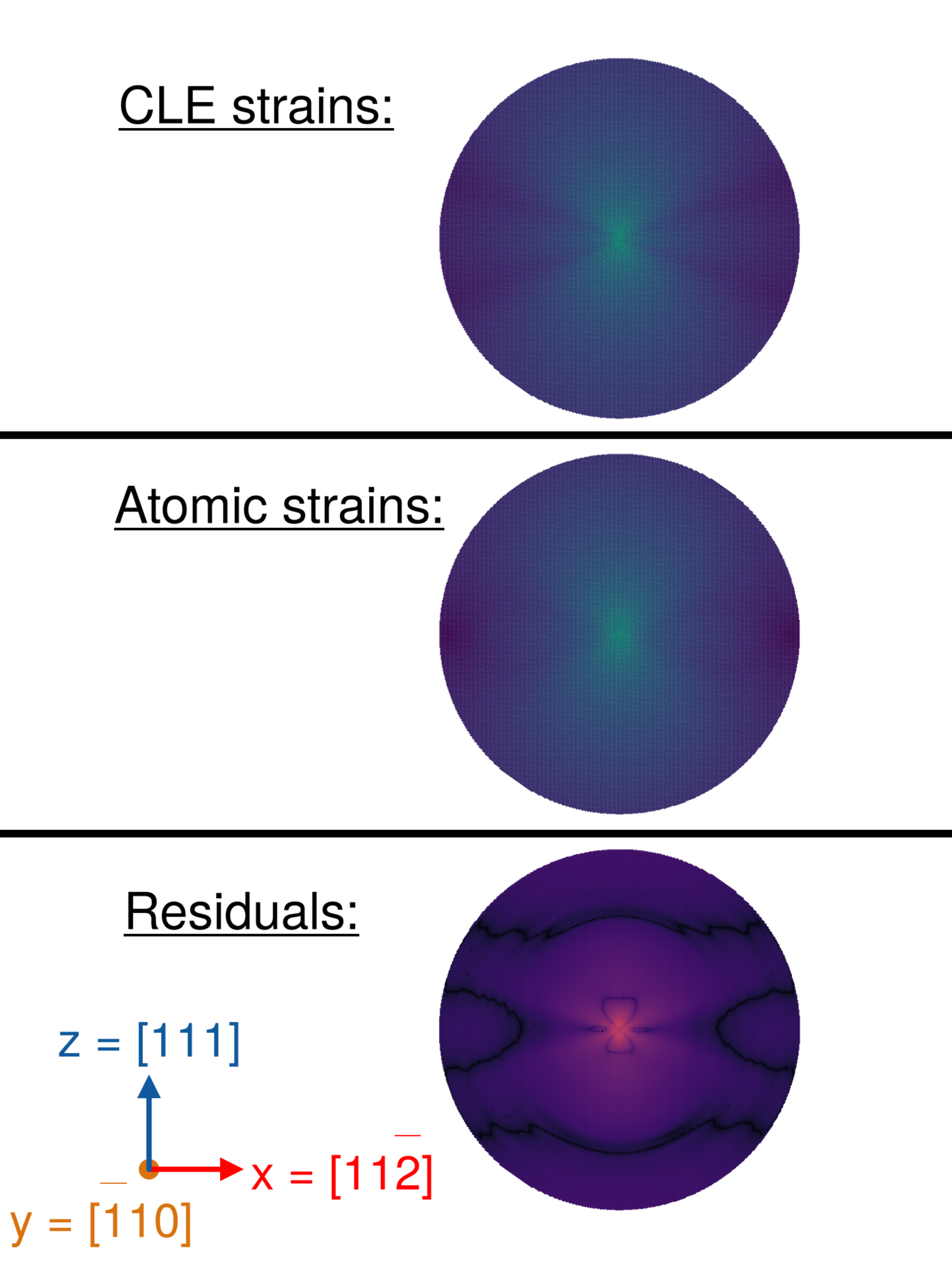}\hspace{-2mm}
	\includegraphics[width=0.33\textwidth]{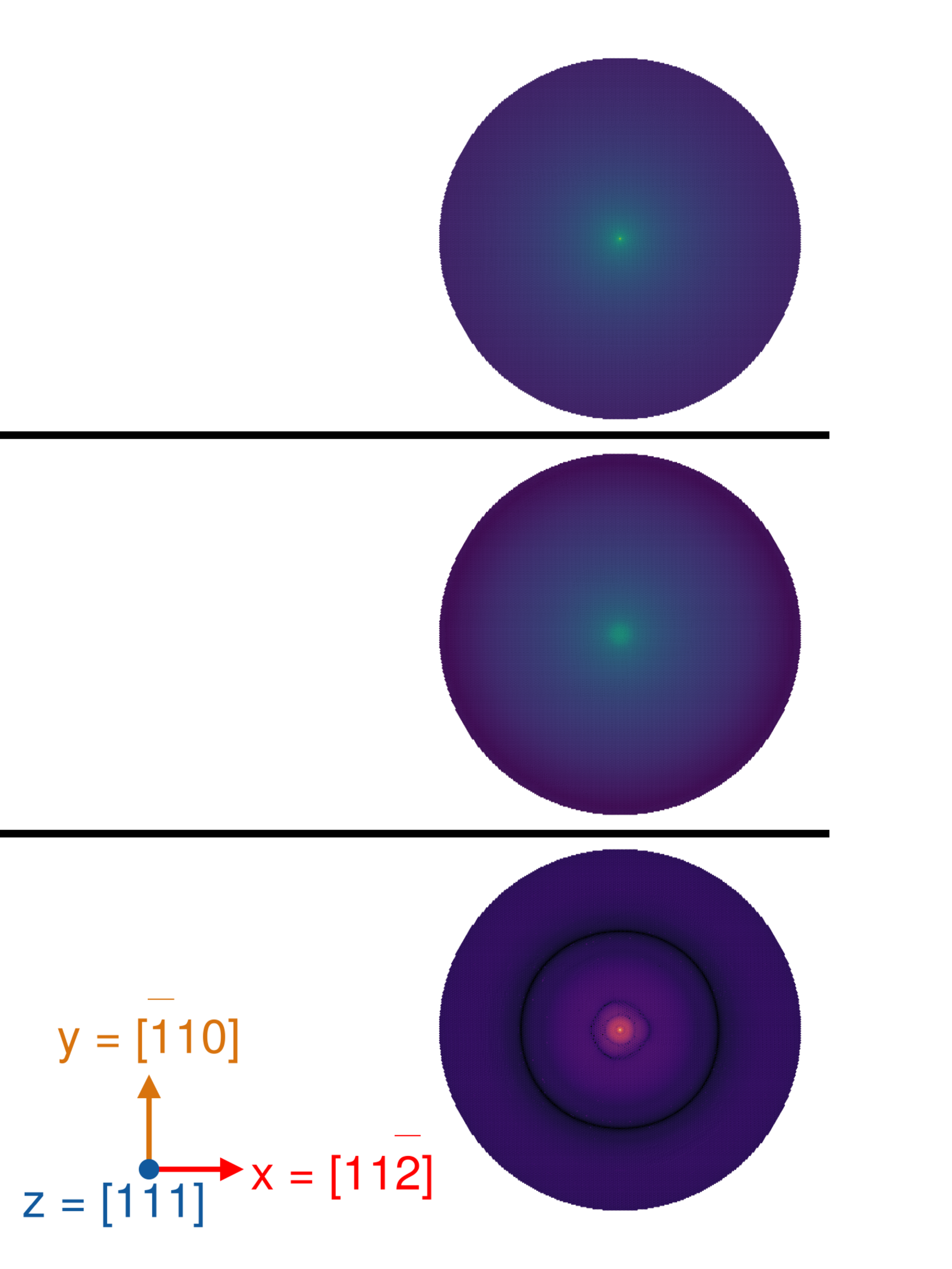}
	\caption{Plots of the shear strain field for both the CLE prediction (top row) and atomistic results (middle row) as calculated using OVITO. The bottom row shows the residuals between the two results including the defect core, and different orientations of the same system are shown in each column, as indicated by the crystalline axes indicated. The strain error is greatest in the vicinity of the dislocation loop, as expected. The system size is a sphere of free atoms with radius of $\sim$ 350 \AA, the dislocation loop has a radius of 12 \AA, and the shell of fixed boundary atoms (not shown in these figures) is around 22 \AA\ thick.}
	\label{fig:strainFields}
\end{figure*}

\Cref{fig:strainFields} shows shear strain fields predicted by both theories at the centre of the simulation cell. The top row of the figure shows the analytic strains, whilst the middle row shows the strain fields calculated from atomistic results. The bottom row shows the residuals between the strains produced by the two results. Each column show the plots sliced along the three co-ordinate axes. The analytical strain fields predicted much larger strains close to the dislocation which was expected when modeling the discrete nature of the problem as a continuum, and this discrepancy is shown by the large residuals between the two datasets near the dislocation core.

Since the dislocation loops studied are prismatic, approximately circular and have normal and Burgers vector oriented in the \(\tfrac12\langle 1 1 1 \rangle \) direction, we expect that the strain fields are axisymmetric about a line passing through the centre of the loop $\mathbf{x}_0$ in the \(\tfrac12\langle 1 1 1 \rangle \) direction. The rightmost column in \cref{fig:strainFields} show this phenomenon, with both the analytic and atomistic simulations demonstrating radial symmetry on the plane containing the dislocation loop.

The residuals of the fields are shown to reduce to around $10^{-4}$ in magnitude, supporting our hypothesis that the analytical model agrees with atomistic simulations in the far-field limit. However, due to surface effects the residuals near the boundary are larger than some regions within the simulation cell, highlighting the importance of comparing the results in a `region of confidence' far enough from the boundary.

\subsubsection{Choice of atomistic boundary condition}
\label{sec:boundary-cond}
The displacement decay rate of the atomistic and analytic simulations was compared for varying simulation cell sizes.
As mentioned at the end of \cref{sec:atomistic-calcs},
one expects that the displacement field behaves linearly proportional to the elastic Green's function \(\sim r^{-2}\) only at the far-field regime.
Therefore, for finite-sized systems one must consider boundary effects as well.
This means that in a finitely sized simulation there is a spherical `shell' of atoms around the dislocation loop where linear elasticity is applicable, and so displacements can only be compared in such a regime for an appropriate verification of results. 

\begin{figure}[!hbpt]
	\includegraphics[width=\linewidth]{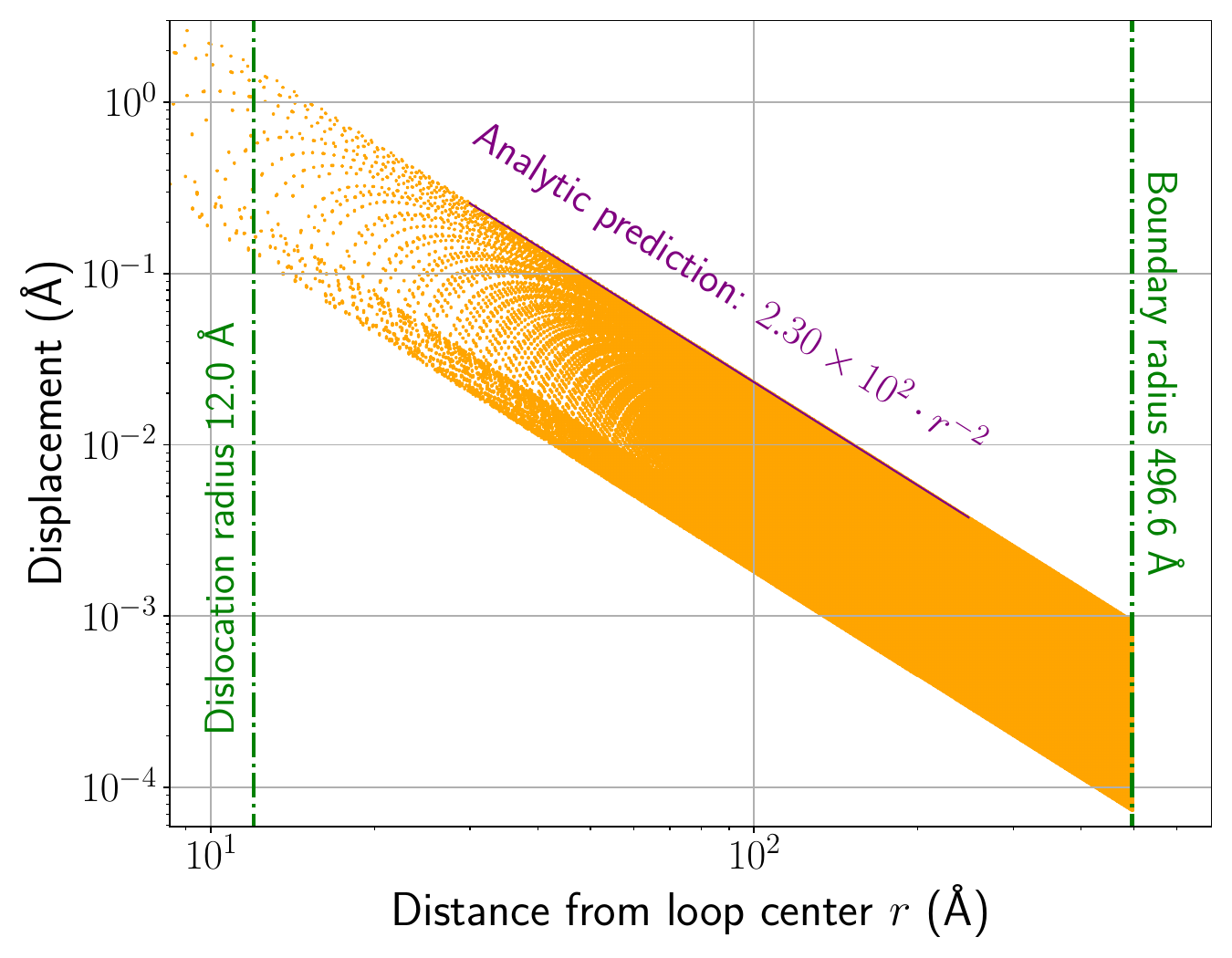}
	\caption{Displacement of atoms computed according to the analytic prediction \cref{eq:leading-order-equation}, which is used for comparison with the atomistic displacements. Each point corresponds to the displacement of a single atomic nucleus.}
	\label{fig:analytic-disps}
\end{figure}

As discussed in \cref{sec:comparison_method}, on order to obtain the largest possible region of comparison, three different choices of boundary condition were studied and the most suitable one was chosen for further analysis. The boundary condition selected was the one which contributed the shortest-range boundary effects on the simulation box. This length-scale was defined as the distance away from the loop centre at which the atomistic results deviated away from the analytic prediction, which determined the `regions of confidence' where the data could be compared. In particular, linear regression was only performed on the data within such regions in order to find the rate of displacement decay predicted by the simulations in the linear regime.
\Cref{fig:analytic-disps} shows the displacement decay of the atoms in the simulation sphere as predicted by the CLE model. The decay rate of the displacement can be seen to be $\sim r^{-2}$ as expected, and this result was used for comparison against the atomistic relaxations for the different boundary conditions considered. 

Three different boundary conditions were tested: first, a fixed boundary condition where the boundary atoms were fixed at the bulk position (denoted as the zero displacement boundary condition), then a fixed boundary conditions with the boundary atoms fixed at the continuum prediction, and finally a traction-free boundary condition. That is, the boundary atoms were free to move during the LAMMPS relaxation of the system. All three cases were studied in order to see which one would contribute the smallest finite size effects on the simulation box, and therefore give the biggest regions inside the simulation sphere where linear elasticity was applicable, which were denoted as the `regions of confidence'. 

\begin{figure*}[htbp!]
	\centering
	\includegraphics[width=0.49\textwidth]{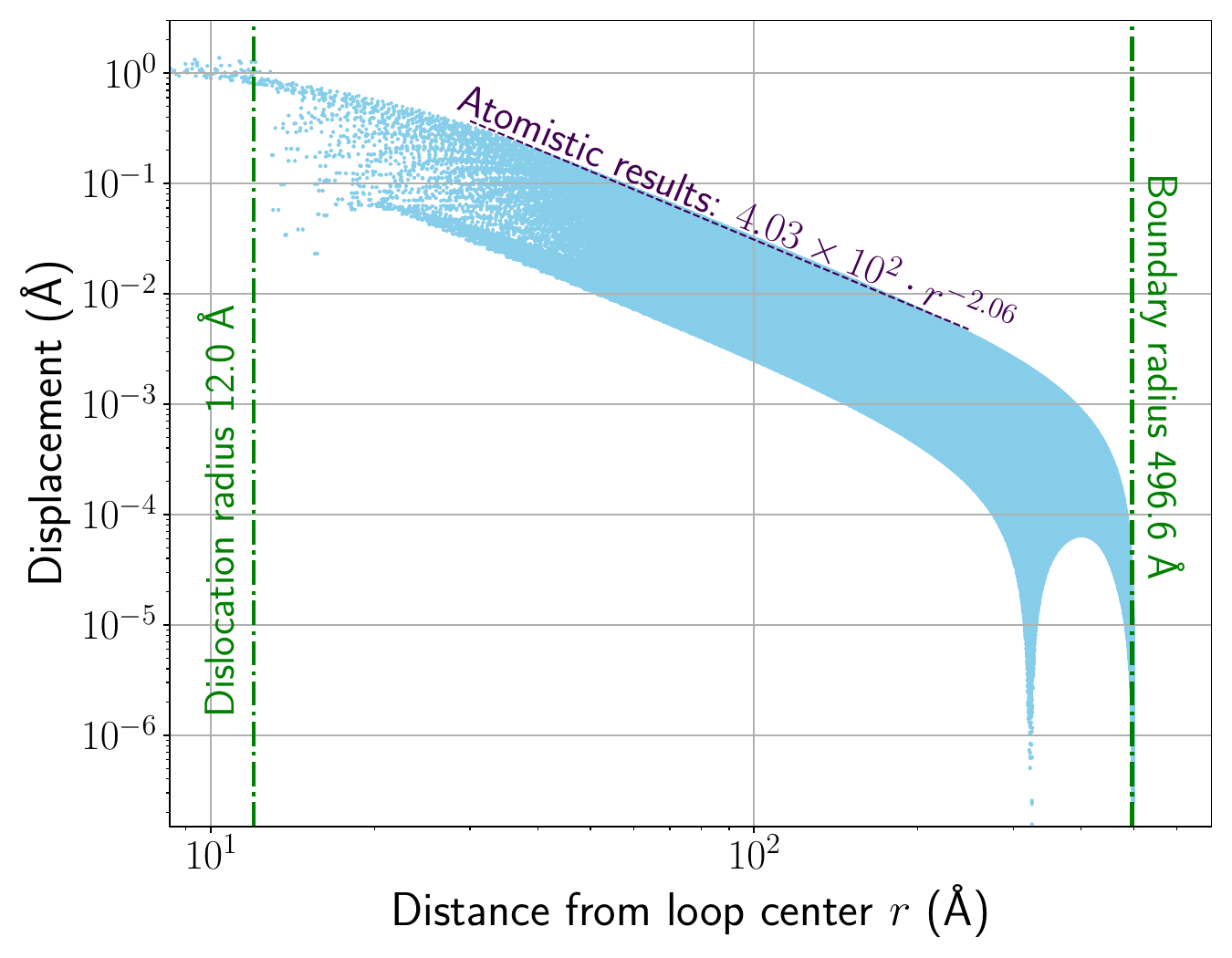}
	\hfill
	\includegraphics[width=0.49\textwidth]{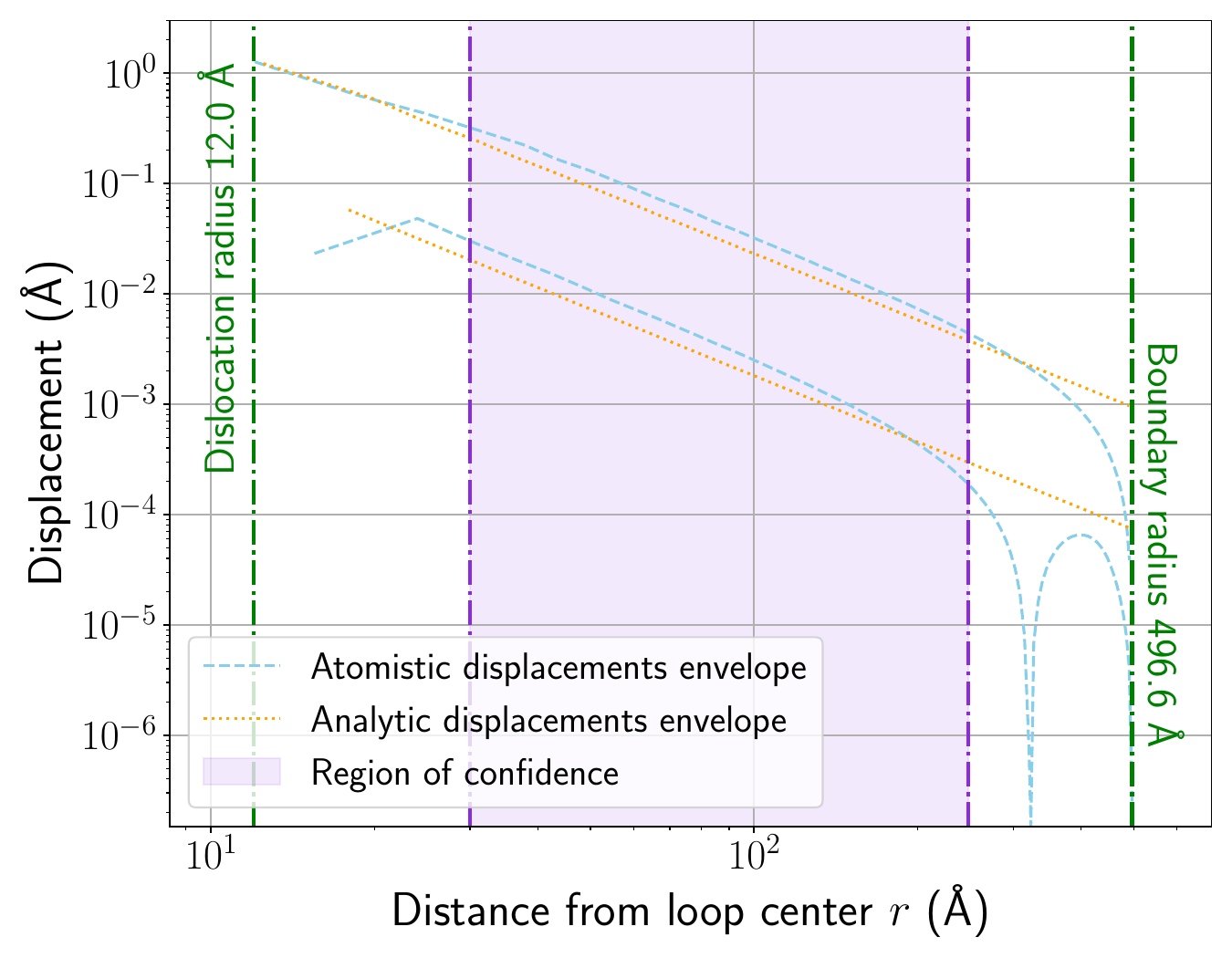}
	\caption{Plots comparing the prediction of the displacements for the fixed boundary condition at zero displacement. Left: Displacement of atoms from molecular statics, with boundary condition fixed at zero displacement; each point reflects the norm of an atomic displacement. Right: The envelopes of the analytic and atomistic displacements, for the fixed boundary condition at zero displacement. The norms of the atomistic displacements were binned according to their distance from the centre of the loop, $\mathbf{x}_0$, and the curves reflect the maxima and minima in these bins.}
	
	\label{fig:boundary-comparison}
\end{figure*}

\Cref{fig:boundary-comparison} shows plots of the displacement prediction from the atomistic simulations on the left and a comparison of the displacement envelopes between the atomistic and CLE prediction on the right. As can be seen from the left plot, the atomistic simulations agree with the continuum theory that the rate of decay of the displacements is $\sim r^{-2}$ within the `region of confidence'. The right plot, where the envelopes are compared, shows how such a region was determined. The regions were constructed such that they began and ended where the top of the envelopes behaved algebraically rather than exponentially - which is shown as a linear behaviour in a logarithmic plot. Quantitatively, for the fixed boundary condition at zero displacement, these values were chosen such that:
\begin{align*}
	f_1 \coloneqq &\,\, \text{max}(a, 2.5 R)  \\
	f_2 \coloneqq &\,\, \text{min}(b, \tfrac{1}{2}L)
\end{align*}
where $f_1$ and $f_2$ are the starting and ending points of the regions respectively, and $a$ and $b$ are the two intersections between the top of the analytic and atomistic displacement envelopes. In the above formulae, $R$ is the dislocation loop radius and $L$ is the system boundary radius. The reasoning behind these choices lies in the following argument: when the two envelope maxima first intersect at point $a$, this indicates that the atomistic results approach the far field behaviour of the continuum theory. However, \cref{fig:boundary-comparison} shows that the atomistic displacements are multiplicatively scaled above the analytic displacements within such a `region of confidence'. For this reason, more care was taken in defining the starting points with some guidance from literature. For the $2.5 R$ criterion, elasticity theory suggests that the stress fields from segments at opposite sides of a circular dislocation loop will tend to cancel at distances which are $\approx 2 R$ away from the loop centre \cite[p.~98]{Hull2011}. The results of \cite{jager_elastic_1975} also reinforce the idea that the `double force' (dipole tensor) approximation to the stress fields is only applicable beyond $2R$. The $\tfrac12L$ criterion for the endpoint $f_2$ was chosen by observing the results from multiple simulations and noticing that exponential displacement decay began at around such distances for different simulation sizes $L$. Although the second criterion is a heuristic, the reasoning is consistent with arguments using Saint-Venant's principle which state that boundary-field effects decay with a characteristic length scale of the same order as the simulation size \cite{horgan_saint-venant_1982, horgan_saint-venant_1994}. The `regions of confidence' therefore presented a metric to determine whether the atomistic results could be considered to be within the linear elastic regime for a reasonable comparison. 

A comparison of the three different boundary conditions and the resulting `regions of confidence' that were determined can be seen in the supplementary material \cite{suppmat}. From comparing all three boundary conditions it was found that the fixed boundary condition at zero displacement provided the largest regions of confidence with a rate of decay close to $r^{-2}$ and thus gave the largest quantity of data that could be compared appropriately between the atomistic and continuum results. 
In light of these investigations, for all subsequent results, the fixed boundary condition at zero displacement was used as it was determined to be the most reasonable choice.

\subsubsection{Comparing different tungsten potentials}

After selecting a boundary condition, the analytic model was then compared against a range of interatomic potentials modelling W-W interactions. For convenience, the potentials considered were given shorthand names: YC \cite{chen_new_2018}, AT \cite{ackland_improved_1987}, DND \cite{Derlet2007}, HZ \cite{han_interatomic_2003}, and ZJ \cite{zhou_misfit-energy-increasing_2004}. All of these potentials are embedded atom models fitted to particular observables for a pure W system, such as cohesive energy, lattice constant, elastic constants and formation energies of different defects, which included SIAs in crowdion sites. For this reason, the potentials considered provided a reasonable range of atomistic models to compare the continuum theory against.

\begin{figure*}
	\centering
	\includegraphics[width=0.9\textwidth]{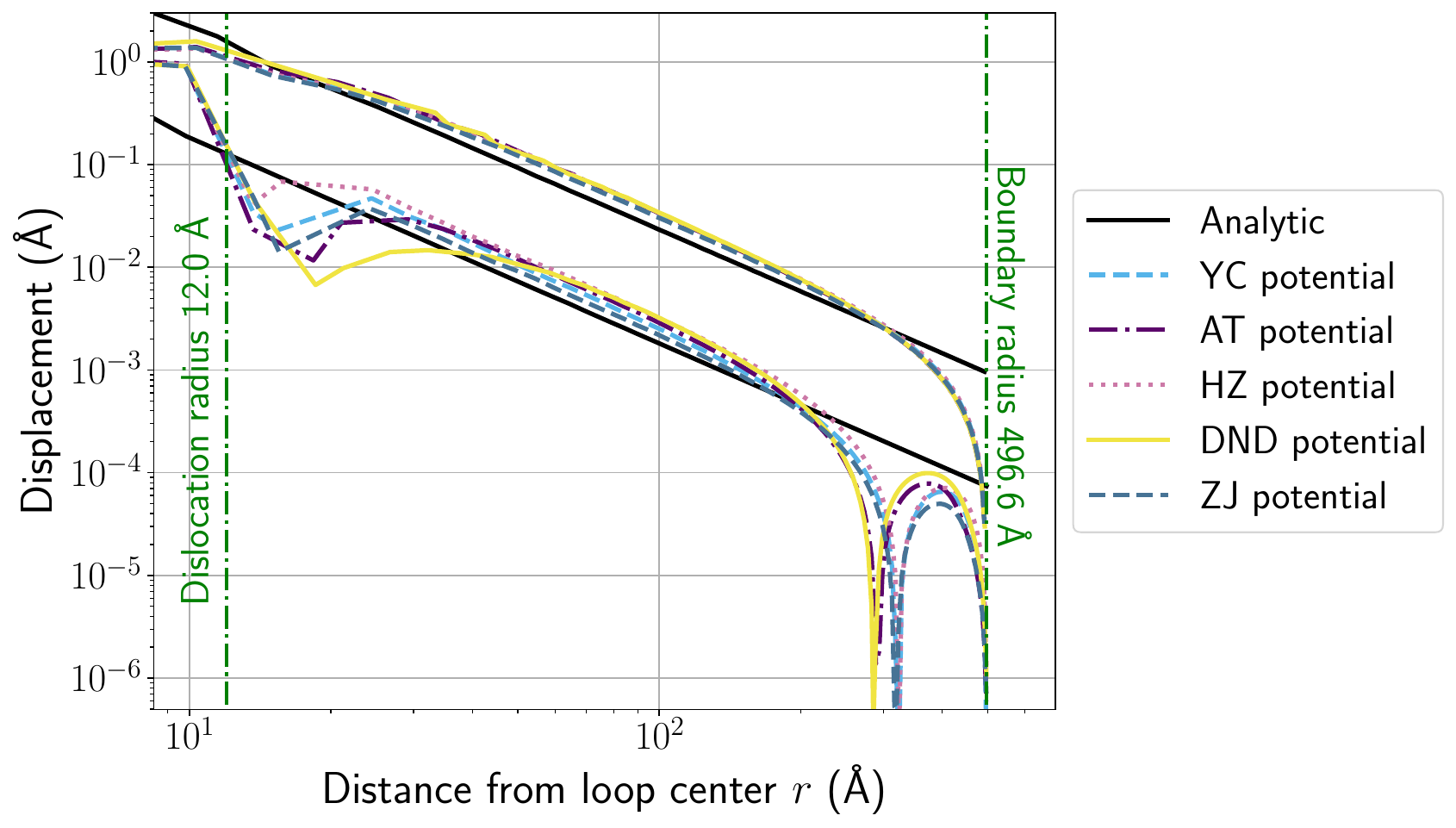}
	\caption{Comparison of displacement envelopes for different potentials against the analytic prediction. Potentials are denoted as YC \cite{chen_new_2018}, AT \cite{ackland_improved_1987}, DND \cite{Derlet2007}, HZ \cite{han_interatomic_2003}, and ZJ \cite{zhou_misfit-energy-increasing_2004}. To obtain the curves, the norms of the atomistic displacements were binned according to their distance from the centre of the loop, and maxima and minima in each bin were computed. At distances far from the dislocation loop, all potentials agree, reflecting the robustness of CLE theory in predicting the far-field displacements.}
	\label{fig:pot-comparison}
\end{figure*}

\Cref{fig:pot-comparison} shows a comparison of the displacement envelopes for all the potentials considered and the analytic prediction computed using \cref{eq:leading-order-equation}. The potentials can be seen to agree similarly with the analytic result in a roughly similar region of confidence, reflecting the robustness of the elastic effects in the far field. Additionally, for all the potentials, the worst rates of displacement decay --- the top of the displacement envelope --- match closely with one another, with more significant variation in the potential's representation of the dislocation core, as expected. From this plot it can be seen that for the chosen boundary condition all of the interatomic potentials agree with the continuum theory to a similar degree, and particularly well for atoms lying at a distance of between 2 and 12 loop radii (24 \AA\ and 144 \AA\ respectively) from $\mathbf{x}_0$. However, all potentials similarly show a multiplicative shift between the atomistic and analytic results within this region, as mentioned before in \Cref{sec:boundary-cond} for the YC \cite{chen_new_2018} potential, which suggests that this shift is due to the finite size effects of the simulation rather than any discrepancies in the atomistic potentials used. This multiplicative shift was studied more in depth later in \cref{sec:loop-radii-variation} where, using convergence arguments, it is demonstrated that the shift can be attributed to finite-size effects, even in the large computational domains used.

The fact that all potentials seem to match clearly can be explained quantitatively by \cref{tab:pot-comp-table}. All potentials used reported similar values of the lattice spacing and elastic coefficients $\C_{11}$, $\C_{12}$ and $\C_{44}$ in W. Additionally, performing a linear fit on the top of the envelopes in the regions confidence within the linear elastic regime yielded rates of decay of $\sim r^{-2}$ for all of the potentials, and prefactors of a similar magnitude. This result verifies that the behaviour of the fields reported is similar to the analytic model in the linear elastic regime.

\begin{table*}[hbpt!]             
	\centering                 
	\renewcommand{\arraystretch}{1.3} 
	
	\begin{ruledtabular}
		\begin{tabular}{cccccccc}
			Potential & Reference & Lattice spacing (\AA) & $\C_{11}$ (GPa) & $\C_{12}$ (GPa) & $\C_{44}$ (GPa) & Decay exponent & Prefactor \\  
			\hline
			YC & \cite{chen_new_2018}  & 3.165 & 522.0 & 204.1 & 161.0 & -2.033 & 283.0  \\ \hline
			AT & \cite{ackland_improved_1987}  & 3.165 & 523.1 & 205.0 & 161.2 & -2.038 & 318.0  \\ \hline
			DND & \cite{Derlet2007}  & 3.165 & 542.4 & 209.0 & 165.1 & -2.036 & 317.0  \\ \hline
			HZ & \cite{han_interatomic_2003}  & 3.165 & 533.5 & 205.9 & 164.1 & -2.038 & 296.7  \\ \hline
			ZJ & \cite{zhou_misfit-energy-increasing_2004}  & 3.165 & 522.5 & 204.2 & 160.8 & -2.059 & 294.2  \\
		\end{tabular}
	\end{ruledtabular}
	\caption{Quantities of interest from the studied potentials; lattice spacings and elastic constants were computed using the \texttt{matscipy} library \cite{Grigorev2024}; all potentials of interest have elastic moduli within a few percent of one another. The latter two columns indicating the fitted rates of decay and prefactors fitted to the data shown in \cref{fig:pot-comparison}; these are found within the regions of confidence described in \cref{sec:boundary-cond}.}     
	\label{tab:pot-comp-table}
\end{table*}

\subsubsection{Effect of loop radius on predicted fields}
\label{sec:loop-radii-variation}
A final study on the atomistic simulations was the effect of varying the dislocation loop radius for fixed simulation sizes, and seeing how this affected the displacement fields predicted by the atomistic theory. 

Initially, the model was derived under the assumption that the area of the dislocation loop was a perfect circle of radius $R$. Using OVITO's dislocation extraction algorithm (DXA) on the relaxed structures \cite{stukowski_automated_2012} a thorough verification of this assumption was performed. DXA finds the dislocation line and the Burgers vectors corresponding to line defects within a structure, and the dislocation line data is stored in array of vertices. Using this array of vertices, an atomistically-informed dislocation loop area can be calculated via the shoelace formula \cite{braden_surveyors_1986}. 
Using this method, it was found that the circular loop was a reasonable assumption for the loop area at least whe
n comparing the visible area enclosed by the dislocation loops from the atomistic simulations.

When the atomistic displacements were compared to the analytic displacements for varying loop radii, however, the results appeared to be shifted by a multiplicative factor. Since the displacement fields are directly proportional to the dislocation loop area as formulated in \cref{eq:leading-order-equation}, this indicated that the relaxation of the structure reported a displacement field corresponding to an effective area significantly different to the one assumed or found by OVITO's DXA. 

\begin{figure*}
	\centering
	\includegraphics[width=3.2in]{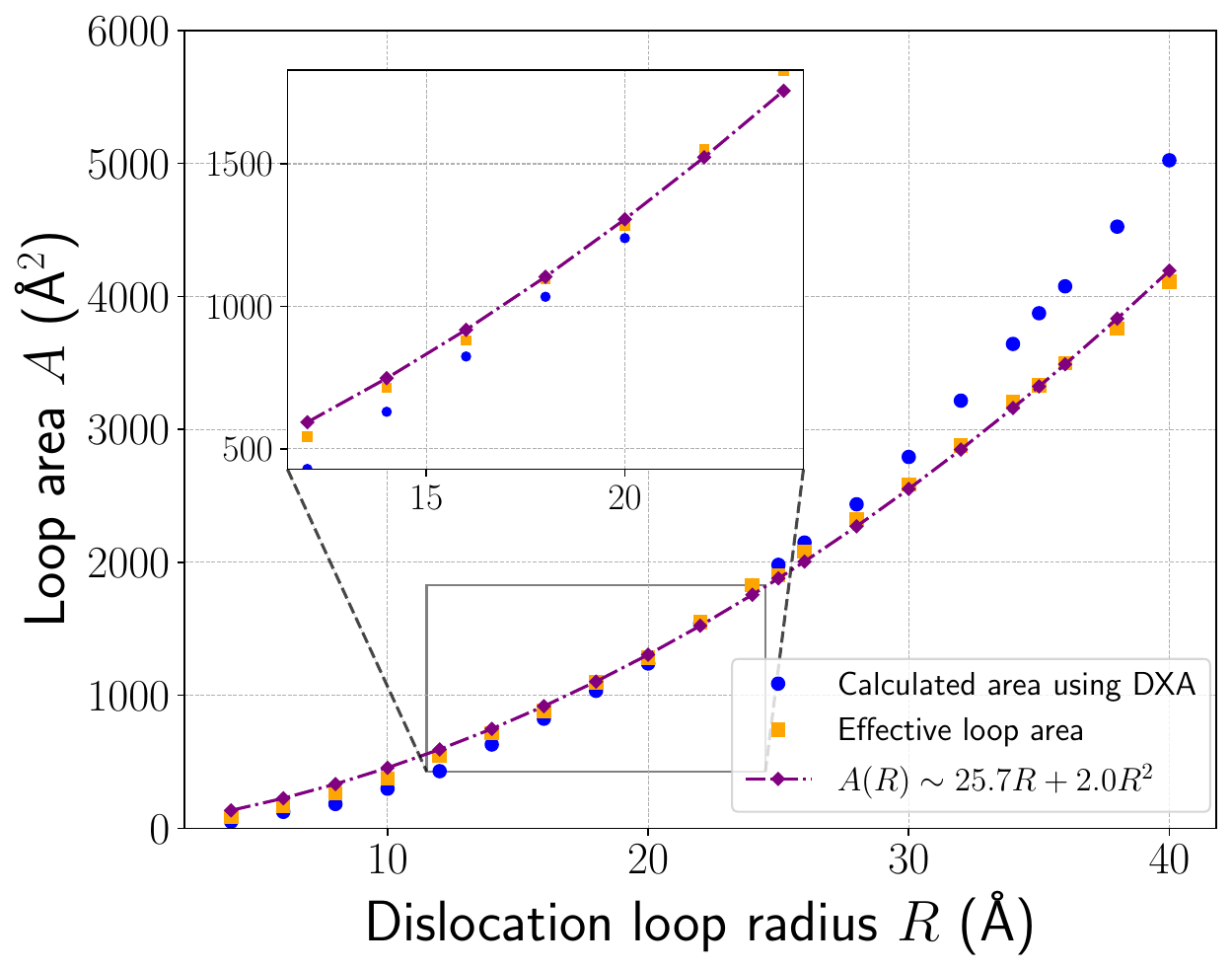}\hfill
	\includegraphics[width=3.2in]{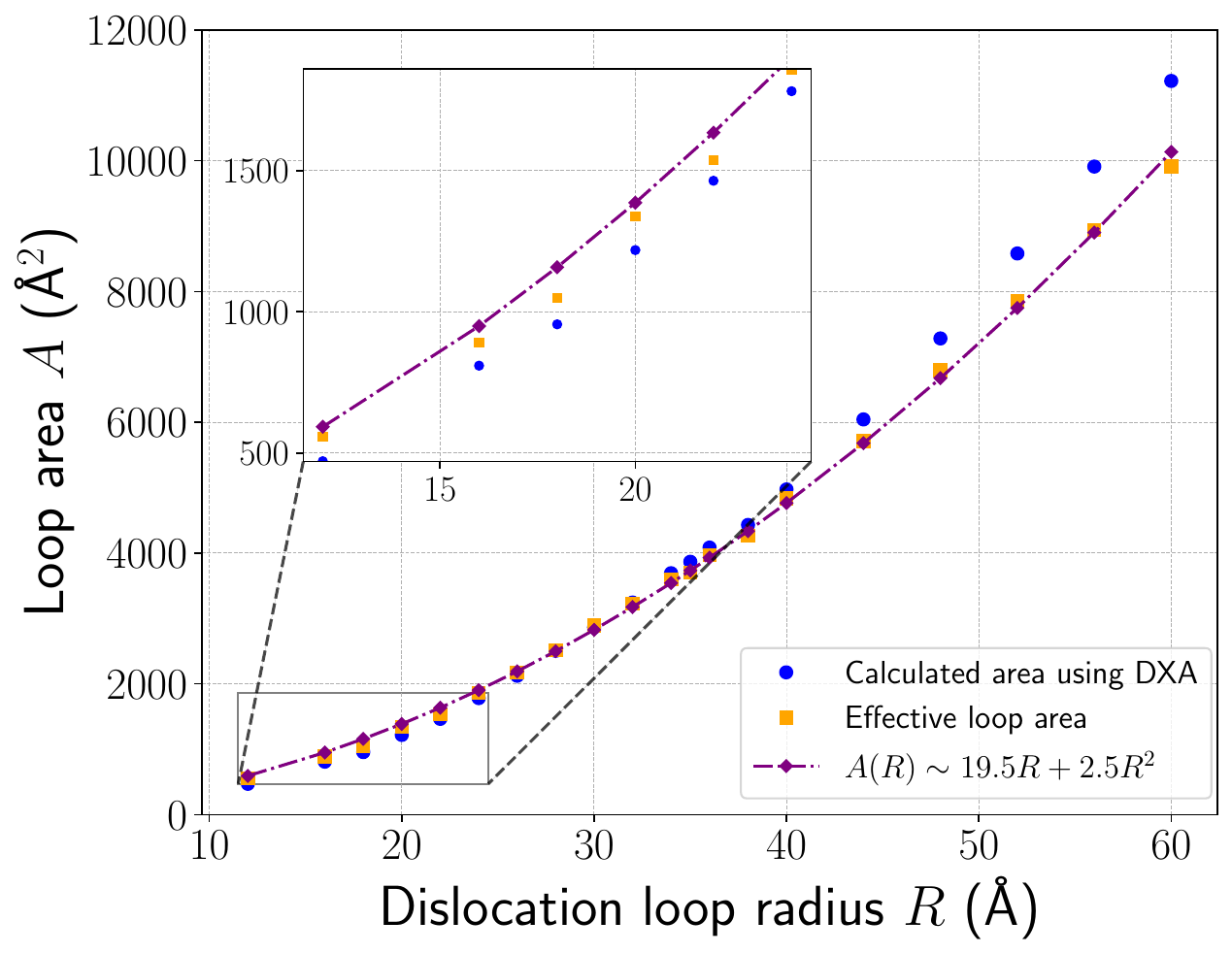}
	
	\caption{Plot of the effective dislocation loop area  obtained from fitting the continuum field to atomistic relaxation data, and a polynomial scaling law $A(R)$ fitted to the data for two different systems sizes. Left: the results for a system size $L=200$ \AA. Right: the results for a system size of $L=500$ \AA. In both cases, the effective areas are compared to the loop area calculated using OVITO's DXA, showing good agreement loop sizes which are small relative to the overall system radius.}
	\label{fig:scalingLaw-200500}
\end{figure*}

\begin{figure}
	\centering
	\includegraphics[width=3.2in]{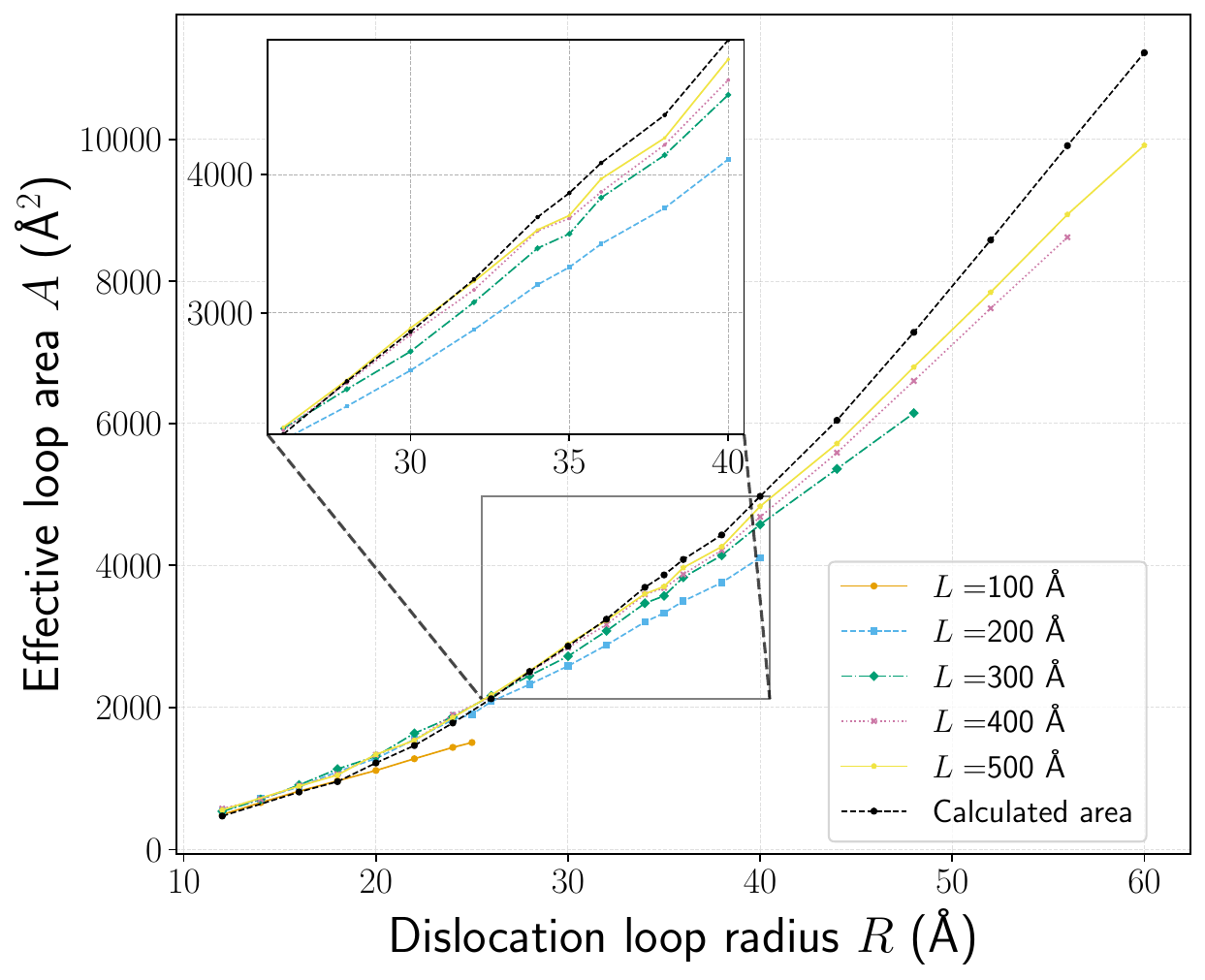}
	\caption{Plot showing the effective area of a dislocation loop for different system sizes (coloured lines), along with the area calculated from the vertices determined by OVITO's dislocation analysis algorithm (black line). As the system size increases, the effective area tends t}
	\label{fig:effective_area_all_systems}
\end{figure}

This motivated a further study in finding a scaling law for the effective area of the dislocation loop for a given loop radius in a finite system. The effective area was found by calculating a scaling factor $c$ between the analytic and atomistic displacements, in a fixed window of radial distance away from the loop. The effective area was then $c \,\times\,A_{\textrm{DXA}}$, for the area $A_{\textrm{DXA}}$ calculated from dislocation analysis. A polynomial scaling law, $A(R) = A_1R + A_2R^2$, was fitted to the effective area data, as a function of the loop radius $R$. The largest order term was assumed to be $R^2$ as the area is expected to have dimensions of length squared. Additionally, a leading constant term was not used, as we expect zero effective loop area for zero magnitude in the radius of the loop. \Cref{fig:scalingLaw-200500} shows a comparison of the fitted scaling law for two different system sizes, one with a simulation sphere with a radius of $L = 200$ \AA\ and the other with a radius of $L = 500$ \AA. As the left panel in \cref{fig:scalingLaw-200500} shows, the linear term in $R$ is more significant for the smaller system size when compared to the larger system size; as the simulation sphere becomes larger, the quadratic term begins to become more significant and the linear term decreases in significance. Analysing the plots in a window of radii, it can be seen that for the smaller system size the effective area crosses the area calculated with DXA at a smaller dislocation loop radius, when compared to the larger system size, indicating that the effective area depends on the total system size. 
\Cref{fig:effective_area_all_systems} supports our conclusion that as the simulation size increases the behaviour of the effective area changes; as might be expected from the asymptotic analysis of \cref{sec:CLEmodel}, to obtain good predictions, the loop radius should be significantly smaller than the overall system size in order to obtain good agreement with the area calculated with DXA.

This analysis raises the question of whether the mismatch in the expected area could be attributed to the finite size effects in the simulation cell and discrete effects from the dislocation core. Recent work in the field of elasticity theory for crystal defects aims to improve such analytic predictions \cite{braun_higher-order_2024} and performing a similar comparison as the one discussed in this paper could demonstrate refinement in the multiplicative shift, although the work needs to be expanded from point defects to dislocation loops. To study the multiplicative shift, further analyses were conducted to see if the effective area was convergent as the simulation size increased, and to predict what value it may be converging to. In order to quantify this convergence, a slice through the data displayed in \cref{fig:effective_area_all_systems} was studied for a fixed radius. In order to quantify the convergence of the effective area, the data was fitted to a power law of the form: $A(L) = A_0(\infty) \, + \, A_1/L^n$, where the parameters $A_0(\infty)$, $A_1$, and $n$ were fitted to. This form was chosen because in the infinite medium limit, we expect that the effective area of the dislocation loop is independent of the system size in such simulations, and is exactly $A_0(\infty)$. For finite system sizes, the second term is significant and therefore this is a way to quantify the effect of the finite system size on the effective area reported. $A_0(\infty), A_1$ and $n$ are all treated as free parameters for the purpose of this fit.

\begin{figure}
	\centering
	\includegraphics[width=3.2in]{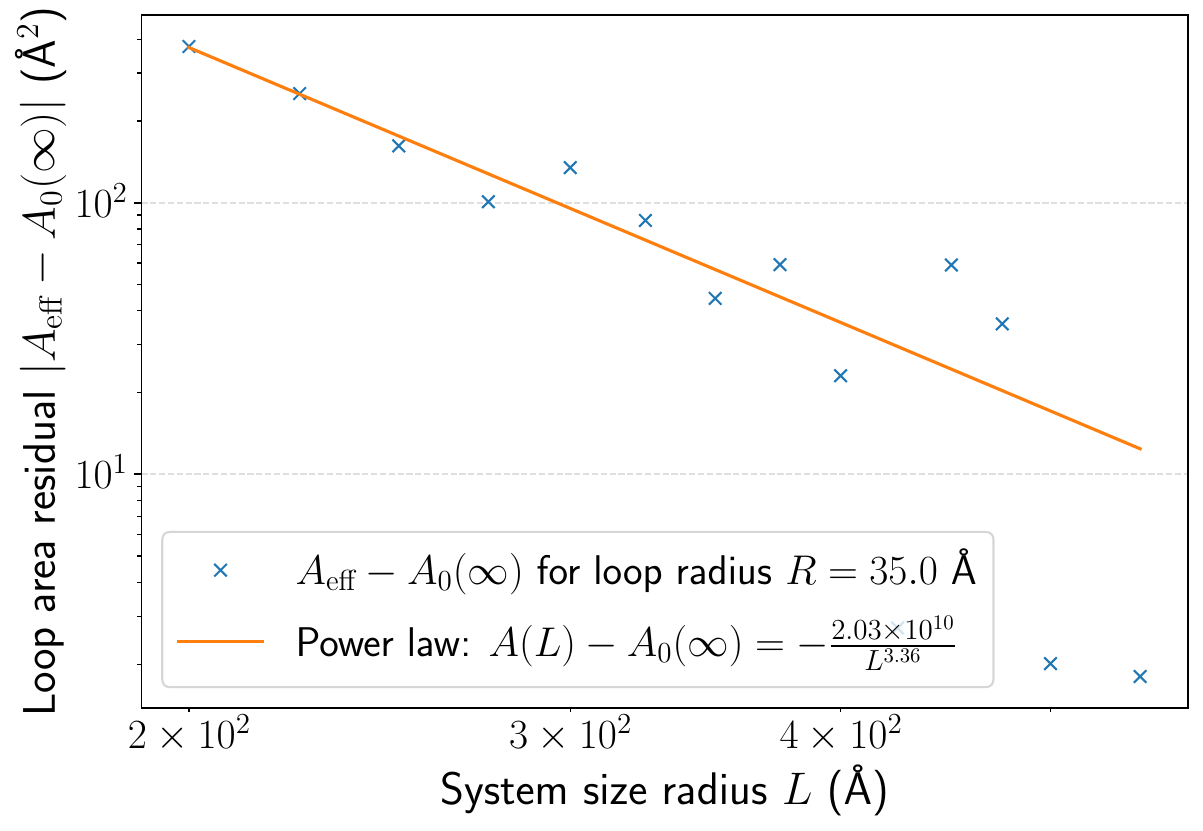}
	\caption{Plot showing the effective area of the dislocation loop as the simulation size increases, rescaled by the fitted parameter $A_0(\infty)$. The area converged asymptotically to the fitted parameter, $A_0(\infty)$, which was reported to be $\sim 3706.4 \,\, \text{\AA}^2$ for a fixed dislocation loop radius of 35 \AA.
	}
	\label{fig:eff_A_conv_R=35.0}
\end{figure}

\Cref{fig:eff_A_conv_R=35.0} shows the convergence of the effective area for a radius of $R = 35$ \AA. The plot shows that the effective area does indeed seem to converge to some value of $A_0(\infty) \approx 3706$ \AA$^2$, at a rate of $\sim L^{-3}$, demonstrating that for small system sizes the finite size effect is very significant, and this effect becomes less important as the data approaches the linear elastic regime of an infinite medium. Additionally, the convergence exponent of $3.36$ is close to a value of 3, which Saint-Venant arguments would predict for a dipole source. The fact this is not exact suggests that there are fitting artifacts in the results, but a study into the dataset shows that as the data is populated with a wider range of system size radii $L$ the convergence exponent approaches a value of 3. Finally, if the value of $A_0(\infty)$ is compared to the area of a circle for a radius of $R = 35$ \AA\ one finds that relative error between $A_0$ and the area of such a circle is around 3.7\%, indicating that in the linear elastic regime dislocation loops can be considered circular within a 3.7\% error margin. Following on from the asymptotic analysis of \cref{sec:CLEmodel}, the error is dependent on the relative size of the dislocation loop with respect to the system size, and thus increases for larger dislocation loops for a fixed system size $L$, which can be seen qualitatively in \cref{fig:effective_area_all_systems}.

\section{Conclusion}
\label{sec:conclusion}
In this paper, a model to represent the fields around dislocation loops using the defect dipole representation was discussed. The continuum model can be applied generally to any dislocation loop for any type of anisotropy due to its dependence on an arbitrary elasticity tensor. The resulting continuum prediction was then compared against atomistic simulations using appropriate interatomic potentials, which verified the comparable behaviour of the displacement and strain fields predicted around the dislocation loops at a qualitative and quantitative level. The anisotropy of the fields is shown to agree between the analytic and atomistic results qualitatively, and the magnitudes are shown to be within similar orders for distances at least twice the loop radii away from the loop centre. 
Additionally, an analysis into the behaviour of the dislocation loop area verified that in the infinite medium, dislocation loops can indeed be considered circular within a relative error margin of around 3.7\%.
Therefore, our results confirm that as the atomistic simulations approached the far-field regime, the continuum model correctly predicts the fields around a dislocation loop within a percentage error, which motivates the use of CLE to describe the interactions between dislocation loops in W.

Further work could include a similar comparison of the predicted fields around voids, another nanoscale defect found in irradiated tungsten. It may be of interest to formulate an elastic correction to the displacement prediction by imposing different boundary conditions, such as Dirichlet fixed boundary conditions at zero displacement, and comparing the result against the atomistic results.
Future plans include the implementation of the Peach-Koehler force in order to perform mesoscale simulations of interactions between dislocation loops and voids due to radiation damage, in order to provide meaningful predictions in the degradation of tungsten for long periods of time under such conditions, and to study any patterning that may arise from such ensembles.

\begin{acknowledgments}
The authors thank Fraser Birks, Matt Nutter and Tom Rocke for their invaluable insight and helpful contributions, along with the entire Warwick Research Software Engineer team for providing crucial help in large scale simulations. JD is supported by the UK Engineering and Physical Sciences Research Council–supported Centre for Doctoral Training in Modelling of Heterogeneous Systems, Grant No. EP/S022848/1, and by the UK Atomic Energy Agency through the funding agreement reference No. 2068257. High-performance computing resources used in this work were provided by the University of Warwick Scientific Computing Research Technology Platform.

\end{acknowledgments}

\bibliography{apssamp, zoteroReading}

@article{PWM2020,
  author  = {Ma, P.-W. and Mason, D. R. and Dudarev, S. L.},
  journal = {Physical Review Materials},
  title   = {Multiscale Analysis of Dislocation Loops and Voids in Tungsten},
  year    = {2020},
}

@article{Bacon1980,
  author  = {Bacon, D. J. and Barnett, D. M. and Scattergood, R. O.},
  journal = {Progress in Materials Science},
  volume = {23},
  pages = {51-262},
  title   = {Anisotropic continuum theory of lattice defects.},
  year    = {1980},
}

@article{Blin1955,
  author  = {Blin, J.},
  journal = {Acta Metallurgica},
  volume = {3(2)},
  pages = {199-200},
  title   = {Energie mutuelle de deux dislocations},
  year    = {1955},
}

@Book{Mura1982,
  author    = {Mura, T.},
  publisher = {Kluwer Academic Publishers},
  title     = {Micromechanics of Defects in Solids},
  year      = {1982},
  isbn      = {902473343X},
}

@Article{Derlet2007,
  author    = {Derlet, P. M. and Nguyen-Manh, D. and Dudarev, S. L.},
  journal   = {Physical Review B},
  title     = {Multiscale modeling of crowdion and vacancy defects in body-centered-cubic transition metals},
  year      = {2007},
  issn      = {1550-235X},
  month     = aug,
  number    = {5},
  pages     = {054107},
  volume    = {76},
  doi       = {10.1103/physrevb.76.054107},
  publisher = {American Physical Society (APS)},
}

@Article{Fitzgerald2018,
  author    = {Fitzgerald, S. P.},
  journal   = {Journal of Micromechanics and Molecular Physics},
  title     = {Structure and dynamics of crowdion defects in bcc metals},
  year      = {2018},
  issn      = {2424-9149},
  month     = sep,
  number    = {03n04},
  pages     = {1840003},
  volume    = {03},
  doi       = {10.1142/s2424913018400039},
  publisher = {World Scientific Pub Co Pte Lt},
}

@Article{HjorthLarsen2017,
  author    = {Hjorth Larsen, A. and Jørgen Mortensen, J. and Blomqvist, J. and Castelli, I. E. and Christensen, R. and Dułak, M. and Friis, J. and Groves, M. N. and Hammer, B. and Hargus, C. and Hermes, E. D. and Jennings, P. C. and Bjerre Jensen, P. and Kermode, J. and Kitchin, J. R. and Leonhard Kolsbjerg, E. and Kubal, J. and Kaasbjerg, K. and Lysgaard, S. and Bergmann Maronsson, J. and Maxson, T. and Olsen, T. and Pastewka, L. and Peterson, A. and Rostgaard, C. and Schiøtz, J. and Schütt, O. and Strange, M. and Thygesen, K. S. and Vegge, T. and Vilhelmsen, L. and Walter, M. and Zeng, Z. and Jacobsen, K. W.},
  journal   = {Journal of Physics: Condensed Matter},
  title     = {The atomic simulation environment—a Python library for working with atoms},
  year      = {2017},
  issn      = {1361-648X},
  month     = jun,
  number    = {27},
  pages     = {273002},
  volume    = {29},
  doi       = {10.1088/1361-648x/aa680e},
  publisher = {IOP Publishing},
}

@Article{Plimpton1995,
  author    = {Plimpton, S.},
  journal   = {Journal of Computational Physics},
  title     = {Fast Parallel Algorithms for Short-Range Molecular Dynamics},
  year      = {1995},
  issn      = {0021-9991},
  month     = mar,
  number    = {1},
  pages     = {1--19},
  volume    = {117},
  doi       = {10.1006/jcph.1995.1039},
  publisher = {Elsevier BV},
}

@Book{Hull2011,
  editor    = {Hull, D. and Bacon, D. J.},
  publisher = {Butterworth-Heinemann},
  title     = {Introduction to dislocations},
  year      = {2011},
  address   = {Oxford},
  edition   = {5th ed.},
  isbn      = {008096673X},
  number    = {v. 37},
  series    = {Materials science and technology (New York, N.Y.)},
  ppn_gvk   = {165500011X},
}

@Misc{jax2018github,
  author = {Bradbury, J. and Frostig, R. and Hawkins, P. and Johnson, M. J. and Leary, C. and Maclaurin, D. and Necula, G. and Paszke, A. and Vander{P}las, J. and Wanderman-{M}ilne, S. and Zhang, Q.},
  title = {{JAX}: composable transformations of {P}ython+{N}um{P}y programs},
  url = {http://github.com/jax-ml/jax},
  version = {0.3.13},
  year = {2018},
}

@article{ovito,
author = {Stukowski, A.},
title = {{Visualization and analysis of atomistic simulation data with OVITO - the
   Open Visualization Tool}},
journal = {Modelling and simulation in materials science and engineering},
year = {{2010}},
volume = {{18}},
number = {{1}},
month = {{Jan}},
}

@Article{Becker2013,
  author    = {Becker, C. A. and Tavazza, F. and Trautt, Z. T. and Buarque de Macedo, R. A.},
  journal   = {Current Opinion in Solid State and Materials Science},
  title     = {Considerations for choosing and using force fields and interatomic potentials in materials science and engineering},
  year      = {2013},
  issn      = {1359-0286},
  month     = dec,
  number    = {6},
  pages     = {277--283},
  volume    = {17},
  doi       = {10.1016/j.cossms.2013.10.001},
  publisher = {Elsevier BV},
}

@Article{Hale2018,
  author    = {Hale, L. M. and Trautt, Z. T. and Becker, C. A.},
  journal   = {Modelling and Simulation in Materials Science and Engineering},
  title     = {Evaluating variability with atomistic simulations: the effect of potential and calculation methodology on the modeling of lattice and elastic constants},
  year      = {2018},
  issn      = {1361-651X},
  month     = may,
  number    = {5},
  pages     = {055003},
  volume    = {26},
  doi       = {10.1088/1361-651x/aabc05},
  publisher = {IOP Publishing},
}

@Misc{LucasHale2016,
  author    = {Hale, L.},
  title     = {NIST Interatomic Potentials Repository},
  year      = {2016},
  copyright = {License Information for NIST data},
  doi       = {10.18434/M37},
  publisher = {National Institute of Standards and Technology},
}

@article{Grigorev2024,
  author = {Grigorev, P. and Frérot, L. and Birks, F. and Gola, A. and Golebiowski, J. and Grießer, J. and Hörmann, J. L. and Klemenz, A. and Moras, G. and Nöhring, W. G. and Oldenstaedt, J. A. and Patel, P. and Reichenbach, T. and Rocke, T. and Shenoy, L. and Walter, M. and Wengert, S. and Zhang, L. and Kermode, J. R. and Pastewka, L.},
  doi = {10.21105/joss.05668},
  journal = {Journal of Open Source Software},
  month = jan,
  number = {93},
  pages = {5668},
  title = {{matscipy: materials science at the atomic scale with Python}},
  url = {https://joss.theoj.org/papers/10.21105/joss.05668},
  volume = {9},
  year = {2024}
}

@Misc{OvitoAtomicStrain,
  howpublished = {\url{https://www.ovito.org/manual/reference/pipelines/modifiers/atomic_strain.html}},
  title        = {Atomic Strain - OVITO User Manual 3.11.2},
}

@Book{Bulatov2020,
  author    = {Bulatov, V. V. and Cai, W.},
  publisher = {Oxford University Press},
  title     = {Computer simulations of dislocations},
  year      = {2020},
  address   = {Oxford},
  isbn      = {9780191916618},
  pagetotal = {1284},
  ppn_gvk   = {1823836054},
}

@Book{Anderson2017,
  author    = {Peter Martin Anderson and John Price Hirth and Jens Lothe},
  publisher = {Cambridge University Press},
  title     = {Theory of dislocations},
  year      = {2017},
  address   = {New York, NY, USA},
  edition   = {Third edition},
  isbn      = {9780521864367},
  pagetotal = {699},
  ppn_gvk   = {860217760},
}

@Misc{suppmat,
  note = {See Supplemental Material at [URL will be inserted by publisher]
          for a comparison of the different boundary conditions considered in atomistic simulations.}
}

@Misc{repo,
	author = {Duque, Joseph and Dudarev, Sergei and Kermode, James and Hudson, Thomas},
	title = {Code for: "Bridging Atomistic and Continuum Descriptions of Nanoscale Dislocation Loops in Tungsten"}, 
	year = {2026},
	doi = {https://doi.org/10.5281/zenodo.18925452},
}

@article{fu_molecular_2019,
	title = {Molecular dynamics simulations of high-energy radiation damage in {W} and {W}–{Re} alloys},
	volume = {524},
	issn = {0022-3115},
	url = {https://www.sciencedirect.com/science/article/pii/S0022311519304416},
	doi = {10.1016/j.jnucmat.2019.06.027},
	urldate = {2026-02-19},
	journal = {Journal of Nuclear Materials},
	author = {Fu, J. and Chen, Y. and Fang, J. and Gao, N. and Hu, W. and Jiang, C. and Zhou, H.B. and Lu, G.H. and Gao, F. and Deng, H.},
	month = oct,
	year = {2019},
	pages = {9--20},
}

@article{clouet_dislocation_2009,
	title = {Dislocation {Core} {Energies} and {Core} {Fields} from {First} {Principles}},
	volume = {102},
	url = {https://link.aps.org/doi/10.1103/PhysRevLett.102.055502},
	doi = {10.1103/PhysRevLett.102.055502},
	number = {5},
	urldate = {2026-02-18},
	journal = {Physical Review Letters},
	publisher = {American Physical Society},
	author = {Clouet, E. and Ventelon, L. and Willaime, F.},
	month = feb,
	year = {2009},
	pages = {055502},
}

@article{yin_computing_2012,
	title = {Computing dislocation stress fields in anisotropic elastic media using fast multipole expansions},
	volume = {20},
	issn = {0965-0393},
	url = {https://doi.org/10.1088/0965-0393/20/4/045015},
	doi = {10.1088/0965-0393/20/4/045015},
	number = {4},
	urldate = {2026-02-18},
	journal = {Modelling and Simulation in Materials Science and Engineering},
	publisher = {IOP Publishing},
	author = {Yin, Jie and Barnett, D M and Fitzgerald, S P and Cai, Wei},
	month = may,
	year = {2012},
	pages = {045015},
}

@article{gao_displacement_2015,
	title = {Displacement fields and self-energies of circular and polygonal dislocation loops in homogeneous and layered anisotropic solids},
	volume = {83},
	issn = {0022-5096},
	url = {https://www.sciencedirect.com/science/article/pii/S002250961500157X},
	doi = {10.1016/j.jmps.2015.06.008},
	urldate = {2026-02-17},
	journal = {Journal of the Mechanics and Physics of Solids},
	author = {Gao, Yanfei and Larson, Bennett C.},
	month = oct,
	year = {2015},
	pages = {104--128},
}

@article{horgan_saint-venant_1982,
	title = {Saint-{Venant} {End} {Effects} {In} {Composites}},
	volume = {16},
	doi = {10.1177/002199838201600506},
	journal = {Journal of Composite Materials},
	author = {Horgan, C.},
	month = sep,
	year = {1982},
	pages = {411--422},
}

@article{horgan_saint-venant_1994,
	title = {Saint-{Venant} end effects in composite structures},
	volume = {4},
	issn = {0961-9526},
	url = {https://www.sciencedirect.com/science/article/pii/0961952694900787},
	doi = {10.1016/0961-9526(94)90078-7},
	number = {3},
	urldate = {2026-02-17},
	journal = {Composites Engineering},
	author = {Horgan, C. O. and Simmonds, J. G.},
	month = jan,
	year = {1994},
	pages = {279--286},
}

@article{jager_elastic_1975,
	title = {Elastic interaction of a dislocation loop with a traction-free surface},
	volume = {31},
	copyright = {Copyright © 1975 WILEY-VCH Verlag GmbH \& Co. KGaA},
	issn = {1521-396X},
	url = {https://onlinelibrary.wiley.com/doi/abs/10.1002/pssa.2210310224},
	doi = {10.1002/pssa.2210310224},
	number = {2},
	urldate = {2026-02-17},
	journal = {phys. status solidi A},
	author = {Jäger, W. and Rühle, M. and Wilkens, M.},
	year = {1975},
	pages = {525--533},
}

@article{fu_effect_2017,
	series = {Computer {Simulation} of {Radiation} effects in {Solids} {Proceedings} of the 13 {COSIRES} {Loughborough}, {UK}, {June} 19-24 2016},
	title = {Effect of collision cascades on dislocations in tungsten: {A} molecular dynamics study},
	volume = {393},
	issn = {0168-583X},
	shorttitle = {Effect of collision cascades on dislocations in tungsten},
	url = {https://www.sciencedirect.com/science/article/pii/S0168583X16304505},
	doi = {10.1016/j.nimb.2016.10.028},
	urldate = {2026-02-16},
	journal = {Nuclear Instruments and Methods in Physics Research Section B: Beam Interactions with Materials and Atoms},
	author = {Fu, B. Q. and Fitzgerald, S. P. and Hou, Q. and Wang, J. and Li, M.},
	month = feb,
	year = {2017},
	pages = {169--173},
}

@article{fitzgerald_shape_2009,
	title = {Shape of prismatic dislocation loops in anisotropic α-{Fe}},
	volume = {89},
	issn = {0950-0839},
	url = {https://doi.org/10.1080/09500830903199012},
	doi = {10.1080/09500830903199012},
	number = {9},
	urldate = {2026-02-16},
	journal = {Philosophical Magazine Letters},
	publisher = {Taylor \& Francis},
	author = {Fitzgerald, S. P. and Yao, Z.},
	month = sep,
	year = {2009},
	pages = {581--588},
}

@article{braden_surveyors_1986,
	title = {The {Surveyor}'s {Area} {Formula}},
	volume = {17},
	issn = {0746-8342},
	url = {https://doi.org/10.1080/07468342.1986.11972974},
	doi = {10.1080/07468342.1986.11972974},
	number = {4},
	urldate = {2025-12-10},
	journal = {The College Mathematics Journal},
	publisher = {Taylor \& Francis},
	author = {Braden, Bart},
	month = sep,
	year = {1986},
	pages = {326--337},
}

@article{ackland_improved_1987,
	title = {An improved {N}-body semi-empirical model for body-centred cubic transition metals},
	volume = {56},
	issn = {0141-8610},
	url = {https://doi.org/10.1080/01418618708204464},
	doi = {10.1080/01418618708204464},
	number = {1},
	urldate = {2025-11-26},
	journal = {Philosophical Magazine A},
	publisher = {Taylor \& Francis},
	author = {Ackland, G. J. and Thetford, R.},
	month = jul,
	year = {1987},
	pages = {15--30},
}

@article{barnett_precise_1972,
	title = {The precise evaluation of derivatives of the anisotropic elastic {Green}'s functions},
	volume = {49},
	issn = {1521-3951},
	url = {https://onlinelibrary.wiley.com/doi/abs/10.1002/pssb.2220490238},
	doi = {10.1002/pssb.2220490238},
	number = {2},
	urldate = {2025-05-21},
	journal = {Phys. Status Solidi B},
	author = {Barnett, D. M.},
	year = {1972},
	pages = {741--748},
}

@article{haussermann_elektronenmikroskopische_1972,
	title = {Elektronenmikroskopische {Untersuchung} der {Strahlenschädigung} durch hochenergetische {Goldionen} in den kubisch-raumzentrierten {Metallen} {Molybdan} und {Wolfram}},
	volume = {25},
	issn = {0031-8086},
	url = {https://doi.org/10.1080/14786437208228893},
	doi = {10.1080/14786437208228893},
	number = {3},
	urldate = {2024-12-11},
	journal = {The Philosophical Magazine: A Journal of Theoretical Experimental and Applied Physics},
	publisher = {Taylor \& Francis},
	author = {Häussermann, Von F.},
	month = mar,
	year = {1972},
	pages = {583--598},
}

@article{rau_vacancy_1968,
	title = {Vacancy dislocation loops in irradiated and annealed tungsten.},
	volume = {18},
	url = {https://www.osti.gov/biblio/4827040},
	doi = {10.1080/14786436808227526},
	urldate = {2024-12-11},
	journal = {Phil. Mag.},
	author = {Rau, R. C.},
	month = jan,
	year = {1968},
	pages = {1079--1084},
}

@misc{braun_higher-order_2024,
	title = {Higher-order {Far}-field {Boundary} {Conditions} for {Crystalline} {Defects}},
	url = {http://arxiv.org/abs/2210.05573},
	doi = {https://doi.org/10.1137/24M165836X},
	urldate = {2024-11-18},
	publisher = {arXiv},
	author = {Braun, Julian and Ortner, Christoph and Wang, Yangshuai and Zhang, Lei},
	month = mar,
	year = {2024},
}

@article{sand_cascade_2017,
	title = {Cascade fragmentation: deviation from power law in primary radiation damage},
	volume = {5},
	shorttitle = {Cascade fragmentation},
	url = {https://doi.org/10.1080/21663831.2017.1294117},
	doi = {10.1080/21663831.2017.1294117},
	number = {5},
	urldate = {2026-02-04},
	journal = {Materials Research Letters},
	publisher = {Taylor \& Francis},
	author = {Sand, A. E. and Mason, D. R. and De Backer, A. and Yi, X. and Dudarev, S. L. and Nordlund, K.},
	month = sep,
	year = {2017},
	pages = {357--363},
}

@techreport{agency_iter_2002,
	title = {{ITER} {Technical} {Basis}},
	url = {https://www.iaea.org/publications/6492/iter-technical-basis},
	urldate = {2025-01-16},
	institution = {International Atomic Energy Agency},
	author = {Agency, International Atomic Energy},
	year = {2002},
	pages = {1},
}

@article{paneth_mechanism_1950,
	title = {The {Mechanism} of {Self}-{Diffusion} in {Alkali} {Metals}},
	volume = {80},
	copyright = {http://link.aps.org/licenses/aps-default-license},
	issn = {0031-899X},
	url = {https://link.aps.org/doi/10.1103/PhysRev.80.708},
	doi = {10.1103/PhysRev.80.708},
	number = {4},
	urldate = {2025-04-23},
	journal = {Physical Review},
	author = {Paneth, Heinz R.},
	month = nov,
	year = {1950},
	pages = {708--711},
}

@article{fitzgerald_structure_2018,
	title = {Structure and dynamics of crowdion defects in bcc metals},
	volume = {03},
	issn = {2424-9130, 2424-9149},
	url = {https://www.worldscientific.com/doi/abs/10.1142/S2424913018400039},
	doi = {10.1142/S2424913018400039},
	number = {03n04},
	urldate = {2024-11-28},
	journal = {Journal of Micromechanics and Molecular Physics},
	author = {Fitzgerald, S. P.},
	month = sep,
	year = {2018},
	pages = {1840003},
}

@article{das_recent_2019,
	title = {Recent advances in characterising irradiation damage in tungsten for fusion power},
	volume = {1},
	issn = {2523-3971},
	url = {https://doi.org/10.1007/s42452-019-1591-0},
	doi = {10.1007/s42452-019-1591-0},
	number = {12},
	urldate = {2025-01-16},
	journal = {SN Applied Sciences},
	author = {Das, Suchandrima},
	month = nov,
	year = {2019},
	pages = {1614},
}

@article{fellman_radiation_2019,
	title = {Radiation damage in tungsten from cascade overlap with voids and vacancy clusters},
	volume = {31},
	issn = {0953-8984},
	url = {https://dx.doi.org/10.1088/1361-648X/ab2ea4},
	doi = {10.1088/1361-648X/ab2ea4},
	number = {40},
	urldate = {2025-01-21},
	journal = {Journal of Physics: Condensed Matter},
	publisher = {IOP Publishing},
	author = {Fellman, A and Sand, A E and Byggmästar, J and Nordlund, K},
	month = jul,
	year = {2019},
	pages = {405402},
}

@article{gilbert_neutron-induced_2011,
	title = {Neutron-induced transmutation effects in {W} and {W}-alloys in a fusion environment},
	volume = {51},
	issn = {0029-5515},
	url = {https://dx.doi.org/10.1088/0029-5515/51/4/043005},
	doi = {10.1088/0029-5515/51/4/043005},
	number = {4},
	urldate = {2025-01-15},
	journal = {Nuclear Fusion},
	author = {Gilbert, M. R. and Sublet, J.-Ch.},
	month = mar,
	year = {2011},
	pages = {043005},
}

@article{de_backer_multiscale_2018,
	title = {Multiscale modelling of the interaction of hydrogen with interstitial defects and dislocations in {BCC} tungsten},
	volume = {58},
	issn = {0029-5515, 1741-4326},
	url = {https://iopscience.iop.org/article/10.1088/1741-4326/aa8e0c},
	doi = {10.1088/1741-4326/aa8e0c},
	number = {1},
	urldate = {2023-01-27},
	journal = {Nuclear Fusion},
	author = {De Backer, A. and Mason, D.R. and Domain, C. and Nguyen-Manh, D. and Marinica, M.-C. and Ventelon, L. and Becquart, C.S. and Dudarev, S.L.},
	month = jan,
	year = {2018},
	pages = {016006},
}

@misc{li_least-square_nodate,
	title = {Least-{Square} {Atomic} {Strain}},
	author = {Li, Ju and Shimizu, Futoshi},
}

@article{hu_irradiation_2016,
	title = {Irradiation hardening of pure tungsten exposed to neutron irradiation},
	volume = {480},
	issn = {00223115},
	url = {https://linkinghub.elsevier.com/retrieve/pii/S0022311516303543},
	doi = {10.1016/j.jnucmat.2016.08.024},
	urldate = {2025-01-16},
	journal = {Journal of Nuclear Materials},
	author = {Hu, X. and Koyanagi, T. and Fukuda, M. and Kumar, N.A.P. K. and Snead, L. L. and Wirth, B. D. and Katoh, Y.},
	month = nov,
	year = {2016},
	pages = {235--243},
}

@article{sand_high-energy_2013,
	title = {High-energy collision cascades in tungsten: {Dislocation} loops structure and clustering scaling laws},
	volume = {103},
	issn = {0295-5075},
	shorttitle = {High-energy collision cascades in tungsten},
	url = {https://dx.doi.org/10.1209/0295-5075/103/46003},
	doi = {10.1209/0295-5075/103/46003},
	number = {4},
	urldate = {2025-01-16},
	journal = {Europhysics Letters},
	publisher = {EDP Sciences, IOP Publishing and Società Italiana di Fisica},
	author = {Sand, A. E. and Dudarev, S. L. and Nordlund, K.},
	month = sep,
	year = {2013},
	pages = {46003},
}

@article{mei_elastic_2023,
	title = {Elastic anisotropy and its temperature dependence for cubic crystals revealed by molecular dynamics simulations},
	volume = {31},
	issn = {0965-0393},
	url = {https://dx.doi.org/10.1088/1361-651X/ace541},
	doi = {10.1088/1361-651X/ace541},
	number = {6},
	urldate = {2025-06-10},
	journal = {Modelling and Simulation in Materials Science and Engineering},
	publisher = {IOP Publishing},
	author = {Mei, Haojie and Wang, Feifei and Li, Jinfu and Kong, Lingti},
	month = jul,
	year = {2023},
	pages = {065013},
}

@incollection{uberuaga_computational_2020,
	address = {Cham},
	title = {Computational {Methods} for {Long}-{Timescale} {Atomistic} {Simulations}},
	isbn = {978-3-319-44677-6},
	url = {https://doi.org/10.1007/978-3-319-44677-6_24},
	doi = {10.1007/978-3-319-44677-6_24},
	urldate = {2025-01-20},
	booktitle = {Handbook of {Materials} {Modeling}: {Methods}: {Theory} and {Modeling}},
	publisher = {Springer International Publishing},
	author = {Uberuaga, Blas Pedro and Perez, Danny},
	editor = {Andreoni, Wanda and Yip, Sidney},
	year = {2020},
	pages = {683--688},
}

@article{stukowski_automated_2012,
	title = {Automated identification and indexing of dislocations in crystal interfaces},
	volume = {20},
	issn = {0965-0393},
	url = {https://doi.org/10.1088/0965-0393/20/8/085007},
	doi = {10.1088/0965-0393/20/8/085007},
	number = {8},
	urldate = {2025-10-21},
	journal = {Modelling and Simulation in Materials Science and Engineering},
	publisher = {IOP Publishing},
	author = {Stukowski, Alexander and Bulatov, Vasily V and Arsenlis, Athanasios},
	month = oct,
	year = {2012},
	pages = {085007},
}

@article{barabash_armour_1999,
	title = {Armour {Materials} for the {ITER} {Plasma} {Facing} {Components}},
	volume = {1999},
	issn = {1402-4896},
	url = {https://iopscience.iop.org/article/10.1238/Physica.Topical.081a00074/meta},
	doi = {10.1238/Physica.Topical.081a00074},
	number = {T81},
	urldate = {2025-01-15},
	journal = {Physica Scripta},
	publisher = {IOP Publishing},
	author = {Barabash, V. and Federici, G. and Matera, R. and Raffray, A. R. and Teams, ITER Home},
	month = jan,
	year = {1999},
	pages = {74},
}

@article{bacon_anisotropic_1980,
	title = {Anisotropic continuum theory of lattice defects},
	volume = {23},
	copyright = {https://www.elsevier.com/tdm/userlicense/1.0/},
	issn = {00796425},
	url = {https://linkinghub.elsevier.com/retrieve/pii/0079642580900079},
	doi = {10.1016/0079-6425(80)90007-9},
	urldate = {2024-11-18},
	journal = {Progress in Materials Science},
	author = {Bacon, D.J. and Barnett, D.M. and Scattergood, R.O.},
	month = jan,
	year = {1980},
	pages = {51--262},
}

@article{swinburne_fast_2016,
	title = {Fast, vacancy-free climb of prismatic dislocation loops in bcc metals},
	volume = {6},
	copyright = {2016 The Author(s)},
	issn = {2045-2322},
	url = {https://www.nature.com/articles/srep30596},
	doi = {10.1038/srep30596},
	number = {1},
	urldate = {2025-01-20},
	journal = {Scientific Reports},
	publisher = {Nature Publishing Group},
	author = {Swinburne, Thomas D. and Arakawa, Kazuto and Mori, Hirotaro and Yasuda, Hidehiro and Isshiki, Minoru and Mimura, Kouji and Uchikoshi, Masahito and Dudarev, Sergei L.},
	month = aug,
	year = {2016},
	pages = {30596},
}

@article{barabash_neutron_2000,
	series = {9th {Int}. {Conf}. on {Fusion} {Reactor} {Materials}},
	title = {Neutron irradiation effects on plasma facing materials},
	volume = {283-287},
	issn = {0022-3115},
	url = {https://www.sciencedirect.com/science/article/pii/S0022311500002038},
	doi = {10.1016/S0022-3115(00)00203-8},
	urldate = {2025-01-16},
	journal = {Journal of Nuclear Materials},
	author = {Barabash, V. and Federici, G. and Rödig, M. and Snead, L. L. and Wu, C. H.},
	month = dec,
	year = {2000},
	pages = {138--146},
}

@article{yoshida_review_1999,
	title = {Review of recent works in development and evaluation of high-{Z} plasma facing materials},
	volume = {266-269},
	issn = {0022-3115},
	url = {https://www.sciencedirect.com/science/article/pii/S0022311598008174},
	doi = {10.1016/S0022-3115(98)00817-4},
	urldate = {2025-01-16},
	journal = {Journal of Nuclear Materials},
	author = {Yoshida, N.},
	month = mar,
	year = {1999},
	pages = {197--206},
}

@article{seidel_activation_2004,
	series = {Proceedings of the 11th {International} {Conference} on {Fusion} {Reactor} {Materials} ({ICFRM}-11)},
	title = {Activation experiment with tungsten in fusion peak neutron field},
	volume = {329-333},
	issn = {0022-3115},
	url = {https://www.sciencedirect.com/science/article/pii/S002231150400337X},
	doi = {10.1016/j.jnucmat.2004.04.145},
	urldate = {2025-01-15},
	journal = {Journal of Nuclear Materials},
	author = {Seidel, K. and Eichin, R. and Forrest, R. A. and Freiesleben, H. and Goncharov, S. A. and Kovalchuk, V. D. and Markovskij, D. V. and Maximov, D. V. and Unholzer, S.},
	month = aug,
	year = {2004},
	pages = {1629--1632},
}

@article{lopez-cazalilla_effect_2023,
	title = {Effect of surface morphology on {Tungsten} sputtering yields},
	volume = {216},
	issn = {0927-0256},
	url = {https://www.sciencedirect.com/science/article/pii/S0927025622005870},
	doi = {10.1016/j.commatsci.2022.111876},
	urldate = {2025-01-15},
	journal = {Computational Materials Science},
	author = {Lopez-Cazalilla, A. and Jussila, J. and Nordlund, K. and Granberg, F.},
	month = jan,
	year = {2023},
	pages = {111876},
}

@article{bolt_materials_2004,
	series = {Proceedings of the 11th {International} {Conference} on {Fusion} {Reactor} {Materials} ({ICFRM}-11)},
	title = {Materials for the plasma-facing components of fusion reactors},
	volume = {329-333},
	issn = {0022-3115},
	url = {https://www.sciencedirect.com/science/article/pii/S0022311504001242},
	doi = {10.1016/j.jnucmat.2004.04.005},
	urldate = {2025-01-16},
	journal = {Journal of Nuclear Materials},
	author = {Bolt, H. and Barabash, V. and Krauss, W. and Linke, J. and Neu, R. and Suzuki, S. and Yoshida, N. and {ASDEX Upgrade Team}},
	month = aug,
	year = {2004},
	pages = {66--73},
}

@article{nygren_making_2011,
	series = {Proceedings of {ICFRM}-14},
	title = {Making tungsten work – {ICFRM}-14 session {T26} paper 501},
	volume = {417},
	issn = {0022-3115},
	url = {https://www.sciencedirect.com/science/article/pii/S0022311510011360},
	doi = {10.1016/j.jnucmat.2010.12.289},
	number = {1},
	urldate = {2025-01-15},
	journal = {Journal of Nuclear Materials},
	author = {Nygren, R. E. and Raffray, R. and Whyte, D. and Urickson, M. A. and Baldwin, M. and Snead, L. L.},
	month = oct,
	year = {2011},
	pages = {451--456},
}

@article{sand_radiation_2014,
	series = {Proceedings of the 16th {International} {Conference} on {Fusion} {Reactor} {Materials} ({ICFRM}-16)},
	title = {Radiation damage production in massive cascades initiated by fusion neutrons in tungsten},
	volume = {455},
	issn = {0022-3115},
	url = {https://www.sciencedirect.com/science/article/pii/S002231151400364X},
	doi = {10.1016/j.jnucmat.2014.06.007},
	number = {1},
	urldate = {2026-02-04},
	journal = {Journal of Nuclear Materials},
	author = {Sand, A. E. and Nordlund, K. and Dudarev, S. L.},
	month = dec,
	year = {2014},
	pages = {207--211},
}

@book{n_i_muskhelishvili_basic_1948,
	title = {Some {Basic} {Problems} {Of} {The} {Mathematical} {Theory} {Of} {Elasticity}},
	isbn = {978-94-017-3034-1},
	url = {http://archive.org/details/muskhelishvili-some-basic-problems-of-the-mathematical-theory-of-elasticity},
	urldate = {2025-04-24},
	publisher = {Springer Dordrecht},
	author = {{N. I. Muskhelishvili}},
	year = {1948},
}

@article{zhou_misfit-energy-increasing_2004,
	title = {Misfit-energy-increasing dislocations in vapor-deposited {CoFe}/{NiFe} multilayers},
	volume = {69},
	url = {https://link.aps.org/doi/10.1103/PhysRevB.69.144113},
	doi = {10.1103/PhysRevB.69.144113},
	number = {14},
	urldate = {2025-11-26},
	journal = {Physical Review B},
	publisher = {American Physical Society},
	author = {Zhou, X. W. and Johnson, R. A. and Wadley, H. N. G.},
	month = apr,
	year = {2004},
	pages = {144113},
}

@article{han_interatomic_2003,
	title = {Interatomic potential for vanadium suitable for radiation damage simulations},
	volume = {93},
	issn = {0021-8979},
	url = {https://doi.org/10.1063/1.1555275},
	doi = {10.1063/1.1555275},
	number = {6},
	urldate = {2025-11-26},
	journal = {Journal of Applied Physics},
	author = {Han, Seungwu and Zepeda-Ruiz, Luis A. and Ackland, Graeme J. and Car, Roberto and Srolovitz, David J.},
	month = mar,
	year = {2003},
	pages = {3328--3335},
}

@article{chen_new_2018,
	title = {New interatomic potentials of {W}, {Re} and {W}-{Re} alloy for radiation defects},
	volume = {502},
	issn = {0022-3115},
	url = {https://www.sciencedirect.com/science/article/pii/S0022311517313417},
	doi = {10.1016/j.jnucmat.2018.01.059},
	urldate = {2025-04-23},
	journal = {Journal of Nuclear Materials},
	author = {Chen, Yangchun and Li, Yu-Hao and Gao, Ning and Zhou, Hong-Bo and Hu, Wangyu and Lu, Guang-Hong and Gao, Fei and Deng, Huiqiu},
	month = apr,
	year = {2018},
	pages = {141--153},
}

@article{peach_forces_1950,
	title = {The {Forces} {Exerted} on {Dislocations} and the {Stress} {Fields} {Produced} by {Them}},
	volume = {80},
	url = {https://link.aps.org/doi/10.1103/PhysRev.80.436},
	doi = {10.1103/PhysRev.80.436},
	number = {3},
	urldate = {2025-06-06},
	journal = {Physical Review},
	publisher = {American Physical Society},
	author = {Peach, M. and Koehler, J. S.},
	month = nov,
	year = {1950},
	pages = {436--439},
}

@article{dudarev_elastic_2017,
	title = {Elastic interactions between nano-scale defects in irradiated materials},
	volume = {125},
	issn = {1359-6454},
	url = {https://www.sciencedirect.com/science/article/pii/S1359645416309284},
	doi = {10.1016/j.actamat.2016.11.060},
	urldate = {2025-06-13},
	journal = {Acta Materialia},
	author = {Dudarev, S. L. and Sutton, A. P.},
	month = feb,
	year = {2017},
	pages = {425--430},
}

@article{lowrie_singlecrystal_1967,
	title = {Single‐{Crystal} {Elastic} {Properties} of {Tungsten} from 24° to 1800°{C}},
	volume = {38},
	issn = {0021-8979},
	url = {https://doi.org/10.1063/1.1709158},
	doi = {10.1063/1.1709158},
	number = {11},
	urldate = {2025-06-10},
	journal = {Journal of Applied Physics},
	author = {Lowrie, Robert and Gonas, A. M.},
	month = oct,
	year = {1967},
	pages = {4505--4509},
}

@article{bolef_elastic_1962,
	title = {Elastic {Constants} of {Single}‐{Crystal} {Mo} and {W} between 77° and 500°{K}},
	volume = {33},
	issn = {0021-8979},
	url = {https://doi.org/10.1063/1.1728952},
	doi = {10.1063/1.1728952},
	number = {7},
	urldate = {2025-06-10},
	journal = {Journal of Applied Physics},
	author = {Bolef, D. I. and De Klerk, J.},
	month = jul,
	year = {1962},
	pages = {2311--2314},
}

@book{balluffi_introduction_2012,
	address = {Cambridge},
	title = {Introduction to {Elasticity} {Theory} for {Crystal} {Defects}},
	isbn = {978-1-107-01255-4},
	url = {https://www.cambridge.org/core/books/introduction-to-elasticity-theory-for-crystal-defects/21F46D0F19C7CD211503757E86800903},
	doi = {10.1017/CBO9780511998379},
	urldate = {2025-06-04},
	publisher = {Cambridge University Press},
	author = {Balluffi, R. W.},
	year = {2012},
}

@article{merola_iter_2010,
	series = {Proceedings of the {Ninth} {International} {Symposium} on {Fusion} {Nuclear} {Technology}},
	title = {{ITER} plasma-facing components},
	volume = {85},
	issn = {0920-3796},
	url = {https://www.sciencedirect.com/science/article/pii/S0920379610004060},
	doi = {10.1016/j.fusengdes.2010.09.013},
	number = {10},
	urldate = {2025-06-06},
	journal = {Fusion Engineering and Design},
	author = {Merola, Mario and Loesser, D. and Martin, A. and Chappuis, P. and Mitteau, R. and Komarov, V. and Pitts, R. A. and Gicquel, S. and Barabash, V. and Giancarli, L. and Palmer, J. and Nakahira, M. and Loarte, A. and Campbell, D. and Eaton, R. and Kukushkin, A. and Sugihara, M. and Zhang, F. and Kim, C. S. and Raffray, R. and Ferrand, L. and Yao, D. and Sadakov, S. and Furmanek, A. and Rozov, V. and Hirai, T. and Escourbiac, F. and Jokinen, T. and Calcagno, B. and Mori, S.},
	month = dec,
	year = {2010},
	pages = {2312--2322},
}

@article{wang_dynamic_2023,
	title = {Dynamic equilibrium of displacement damage defects in heavy-ion irradiated tungsten},
	volume = {244},
	issn = {1359-6454},
	url = {https://www.sciencedirect.com/science/article/pii/S1359645422009533},
	doi = {10.1016/j.actamat.2022.118578},
	urldate = {2025-06-04},
	journal = {Acta Materialia},
	author = {Wang, Shiwei and Guo, Wangguo and Schwarz-Selinger, Thomas and Yuan, Yue and Ge, Lin and Cheng, Long and Zhang, Xiaona and Cao, Xingzhong and Fu, Engang and Lu, Guang-Hong},
	month = jan,
	year = {2023},
	pages = {118578},
}

@article{bitzek_structural_2006,
	title = {Structural {Relaxation} {Made} {Simple}},
	volume = {97},
	url = {https://link.aps.org/doi/10.1103/PhysRevLett.97.170201},
	doi = {10.1103/PhysRevLett.97.170201},
	number = {17},
	urldate = {2025-05-28},
	journal = {Physical Review Letters},
	publisher = {American Physical Society},
	author = {Bitzek, Erik and Koskinen, Pekka and Gähler, Franz and Moseler, Michael and Gumbsch, Peter},
	month = oct,
	year = {2006},
	pages = {170201},
}

@article{polyak_conjugate_1969,
	title = {The conjugate gradient method in extremal problems},
	volume = {9},
	issn = {0041-5553},
	url = {https://www.sciencedirect.com/science/article/pii/0041555369900354},
	doi = {10.1016/0041-5553(69)90035-4},
	number = {4},
	urldate = {2025-05-28},
	journal = {USSR Computational Mathematics and Mathematical Physics},
	author = {Polyak, B. T.},
	month = jan,
	year = {1969},
	pages = {94--112},
}

@article{cai_non-singular_2006,
	title = {A non-singular continuum theory of dislocations},
	volume = {54},
	issn = {0022-5096},
	url = {https://www.sciencedirect.com/science/article/pii/S002250960500195X},
	doi = {10.1016/j.jmps.2005.09.005},
	number = {3},
	urldate = {2025-02-03},
	journal = {Journal of the Mechanics and Physics of Solids},
	author = {Cai, Wei and Arsenlis, Athanasios and Weinberger, Christopher R. and Bulatov, Vasily V.},
	month = mar,
	year = {2006},
	pages = {561--587},
}

@article{rovelli_statistical_2018,
	title = {Statistical model for diffusion-mediated recovery of dislocation and point-defect microstructure},
	volume = {98},
	url = {https://link.aps.org/doi/10.1103/PhysRevE.98.043002},
	doi = {10.1103/PhysRevE.98.043002},
	number = {4},
	urldate = {2025-01-21},
	journal = {Physical Review E},
	publisher = {American Physical Society},
	author = {Rovelli, I. and Dudarev, S. L. and Sutton, A. P.},
	month = oct,
	year = {2018},
	pages = {043002},
}

@article{li_diffusion_2019,
	title = {Diffusion and interaction of prismatic dislocation loops simulated by stochastic discrete dislocation dynamics},
	volume = {3},
	url = {https://link.aps.org/doi/10.1103/PhysRevMaterials.3.073805},
	doi = {10.1103/PhysRevMaterials.3.073805},
	number = {7},
	urldate = {2025-01-21},
	journal = {Physical Review Materials},
	publisher = {American Physical Society},
	author = {Li, Yang and Boleininger, Max and Robertson, Christian and Dupuy, Laurent and Dudarev, Sergei L.},
	month = jul,
	year = {2019},
	pages = {073805},
}

@article{toschi_how_2001,
	title = {How far is a fusion power reactor from an experimental reactor},
	volume = {56-57},
	issn = {0920-3796},
	url = {https://www.sciencedirect.com/science/article/pii/S0920379601005774},
	doi = {10.1016/S0920-3796(01)00577-4},
	urldate = {2025-01-21},
	journal = {Fusion Engineering and Design},
	author = {Toschi, R and Barabaschi, P and Campbell, D and Elio, F and Maisonnier, D and Ward, D},
	month = oct,
	year = {2001},
	pages = {163--172},
}

@incollection{ruiz_pestana_chapter_2023,
	title = {Chapter {Two} - {Atomistic} molecular modeling methods},
	isbn = {978-0-12-823021-3},
	url = {https://www.sciencedirect.com/science/article/pii/B9780128230213000063},
	doi = {10.1016/B978-0-12-823021-3.00006-3},
	urldate = {2025-01-21},
	booktitle = {Fundamentals of {Multiscale} {Modeling} of {Structural} {Materials}},
	publisher = {Elsevier},
	author = {Ruiz Pestana, Luis Alberto and Liao, Yangchao and Li, Zhaofan and Xia, Wenjie},
	editor = {Xia, Wenjie and Ruiz Pestana, Luis Alberto},
	month = jan,
	year = {2023},
	pages = {37--73},
}

@incollection{dudarev_chapter_2025,
	title = {Chapter 3.10 - {Modelling} and simulation of fusion materials},
	isbn = {978-0-443-13629-0},
	url = {https://www.sciencedirect.com/science/article/pii/B9780443136290000125},
	doi = {10.1016/B978-0-443-13629-0.00012-5},
	urldate = {2025-01-20},
	booktitle = {Fusion {Energy} {Technology} {R}\&{D} {Priorities}},
	publisher = {Elsevier},
	author = {Dudarev, S. L.},
	editor = {El-Guebaly, Laila A.},
	month = jan,
	year = {2025},
	pages = {93--97},
}

@article{swinburne_theory_2013,
	title = {Theory and simulation of the diffusion of kinks on dislocations in bcc metals},
	volume = {87},
	url = {https://link.aps.org/doi/10.1103/PhysRevB.87.064108},
	doi = {10.1103/PhysRevB.87.064108},
	number = {6},
	urldate = {2025-01-20},
	journal = {Physical Review B},
	publisher = {American Physical Society},
	author = {Swinburne, T. D. and Dudarev, S. L. and Fitzgerald, S. P. and Gilbert, M. R. and Sutton, A. P.},
	month = feb,
	year = {2013},
	pages = {064108},
}

@article{das_dislocation_2024,
	title = {Dislocation pinning in helium-implanted tungsten: {A} molecular dynamics study},
	volume = {601},
	issn = {0022-3115},
	shorttitle = {Dislocation pinning in helium-implanted tungsten},
	url = {https://www.sciencedirect.com/science/article/pii/S0022311524003957},
	doi = {10.1016/j.jnucmat.2024.155293},
	urldate = {2025-01-20},
	journal = {Journal of Nuclear Materials},
	author = {Das, Suchandrima and Sand, Andrea and Hofmann, Felix},
	month = dec,
	year = {2024},
	pages = {155293},
}

@article{xu_molecular_2024,
	title = {Molecular dynamics investigation of dislocation-hydrogen/helium interactions in tungsten},
	volume = {592},
	issn = {0022-3115},
	url = {https://www.sciencedirect.com/science/article/pii/S0022311524000515},
	doi = {10.1016/j.jnucmat.2024.154948},
	urldate = {2025-01-20},
	journal = {Journal of Nuclear Materials},
	author = {Xu, Bai-Chuan and Li, Xiao-Chun and Wang, Jinlong and Li, Ya-Wen and Pan, Xin-Dong and Zhou, Hai-Shan and Luo, Guang-Nan},
	month = apr,
	year = {2024},
	pages = {154948},
}

@article{kobayashi_molecular_2015,
	series = {{PLASMA}-{SURFACE} {INTERACTIONS} 21},
	title = {A molecular dynamics study on bubble growth in tungsten under helium irradiation},
	volume = {463},
	issn = {0022-3115},
	url = {https://www.sciencedirect.com/science/article/pii/S002231151400991X},
	doi = {10.1016/j.jnucmat.2014.12.049},
	urldate = {2025-01-20},
	journal = {Journal of Nuclear Materials},
	author = {Kobayashi, Ryo and Hattori, Tatsunori and Tamura, Tomoyuki and Ogata, Shuji},
	month = aug,
	year = {2015},
	pages = {1071--1074},
}

@article{fikar_molecular_2009,
	series = {Fusion {Reactor} {Materials}},
	title = {Molecular dynamics simulation of radiation damage in bcc tungsten},
	volume = {386-388},
	issn = {0022-3115},
	url = {https://www.sciencedirect.com/science/article/pii/S0022311508008143},
	doi = {10.1016/j.jnucmat.2008.12.068},
	urldate = {2025-01-20},
	journal = {Journal of Nuclear Materials},
	author = {Fikar, J. and Schäublin, R.},
	month = apr,
	year = {2009},
	pages = {97--101},
}

@article{srivastava_dislocation_2013,
	title = {Dislocation motion in tungsten: {Atomistic} input to discrete dislocation simulations},
	volume = {47},
	issn = {0749-6419},
	shorttitle = {Dislocation motion in tungsten},
	url = {https://www.sciencedirect.com/science/article/pii/S0749641913000247},
	doi = {10.1016/j.ijplas.2013.01.014},
	urldate = {2025-01-16},
	journal = {International Journal of Plasticity},
	author = {Srivastava, K. and Gröger, R. and Weygand, D. and Gumbsch, P.},
	month = aug,
	year = {2013},
	pages = {126--142},
}

@article{thompson_damage_1960,
	title = {The damage and recovery of neutron irradiated tungsten},
	volume = {5},
	issn = {0031-8086},
	url = {https://doi.org/10.1080/14786436008235842},
	doi = {10.1080/14786436008235842},
	number = {51},
	urldate = {2025-01-16},
	journal = {The Philosophical Magazine: A Journal of Theoretical Experimental and Applied Physics},
	publisher = {Taylor \& Francis},
	author = {Thompson, M. W.},
	month = mar,
	year = {1960},
	note = {\_eprint: https://doi.org/10.1080/14786436008235842},
	pages = {278--296},
}

@article{kaufmann_tungsten_2007,
	series = {Proceedings of the 24th {Symposium} on {Fusion} {Technology}},
	title = {Tungsten as first wall material in fusion devices},
	volume = {82},
	issn = {0920-3796},
	url = {https://www.sciencedirect.com/science/article/pii/S0920379607001391},
	doi = {10.1016/j.fusengdes.2007.03.045},
	number = {5},
	urldate = {2025-01-15},
	journal = {Fusion Engineering and Design},
	author = {Kaufmann, M. and Neu, R.},
	month = oct,
	year = {2007},
	pages = {521--527},
}

@article{fukuzumi_defect_2005,
	series = {Proceedings of the 6th {International} {Workshop} on {Spallation} {Materials} {Technology}},
	title = {Defect structural evolution in high purity tungsten irradiated with electrons using high voltage electron microscope},
	volume = {343},
	issn = {0022-3115},
	url = {https://www.sciencedirect.com/science/article/pii/S0022311505001601},
	doi = {10.1016/j.jnucmat.2004.11.019},
	number = {1},
	urldate = {2024-12-11},
	journal = {Journal of Nuclear Materials},
	author = {Fukuzumi, S. and Yoshiie, T. and Satoh, Y. and Xu, Q. and Mori, H. and Kawai, M.},
	month = aug,
	year = {2005},
	pages = {308--312},
}

@article{derlet_multiscale_2007,
	title = {Multiscale modeling of crowdion and vacancy defects in body-centered-cubic transition metals},
	volume = {76},
	url = {https://link.aps.org/doi/10.1103/PhysRevB.76.054107},
	doi = {10.1103/PhysRevB.76.054107},
	number = {5},
	urldate = {2024-11-25},
	journal = {Physical Review B},
	publisher = {American Physical Society},
	author = {Derlet, P. M. and Nguyen-Manh, D. and Dudarev, S. L.},
	month = aug,
	year = {2007},
	pages = {054107},
}

\end{document}